\documentclass{aa}  
\usepackage{txfonts}
\usepackage{graphicx}
\usepackage{natbib}
\def\h2s{H$_{2}$S}
\def\oo{ON--OFF}
\def\so2{$\textrm{SO}_{2}$}
\begin{document}

   \title{Interferometric imaging of the sulfur-bearing molecules H$_2$S, SO and CS in comet C/1995 O1 (Hale-Bopp)\thanks{Based on observations carried out with
the IRAM Plateau de Bure Interferometer. IRAM is supported by INSU/CNRS (France), MPG (Germany) and IGN (Spain).}}

   \author{J. Boissier\inst{1}
          \and
          D. Bockel\'ee-Morvan\inst{1}
          \and
          N. Biver\inst{1}
          \and
          J. Crovisier\inst{1}
          \and
          D. Despois\inst{2} 
          \and
          B.G. Marsden\inst{3}
          \and
          R. Moreno\inst{1}
          }

   \offprints{J. Boissier}

   \institute{LESIA, Observatoire de Paris-Meudon,
               5 place Jules Janssen, 92195 Meudon, France\\
              \email{jeremie.boissier@obspm.fr}
    \and                  
              Universit\'e Bordeaux 1, BP 89, 2 rue de l'Observatoire 33270 Floirac, France
    \and 
              Harvard-Smithsonian Center for Astrophysics, Cambridge, MA 02138, USA \\
}

%   \date{Received September 15, 1996; accepted March 16, 1997}

\abstract
{We present observations of rotational lines of \h2s, SO and CS performed  in comet C/1995 O1 (Hale-Bopp) in March 1997 with the Plateau de Bure  interferometer of Institut de Radioastronomie  Millim\'etrique (IRAM).}
{The observations provide informations on the spatial and velocity  distributions of these molecules.
They can be used to constrain their photodissociation rate and their origin in the coma, i.e. nucleus  or parent source. }
{We use a radiative transfer code which allows us to compute synthetic line profiles and interferometric maps, to be compared to the observations.} 
{Both single-dish spectra and interferometric spectral maps show  evidence for a day/night asymmetry in the outgassing.
 From the analysis of the spectral maps, including the astrometry, we show that SO and CS present in addition a jet-like structure that may be the gaseous  counterpart of the dust high-latitude jet observed in optical images. 
A CS rotating jet is also observed.
Using the astrometry provided by continuum radio maps obtained in parallel at IRAM, we conclude that there is no need to invoke the existence of nongravitational forces acting on this comet, and provide an updated orbit.           
The radial extension of \h2s is found to be consistent with direct release of \h2s from the nucleus. SO displays an extended radial distribution. 
Assuming that SO$_2$ is the parent of SO, the photodissociation rate of SO is measured to be $1.5 \times 10^{-4} \, \mathrm{s}^{-1}$ at 1 AU from the Sun. 
This is lower than most laboratory-based estimates and may suggest that SO is not solely produced by SO$_2$ photolysis.
From the observations of $J$(2--1) and $J$(5--4) CS lines, we deduce a  CS photodissociation rate of $1-5 \times 10^{-5} \, \mathrm{s}^{-1}$.
The photodissociation rate of CS$_2$, the likely parent of CS, cannot be constrained due to insufficient angular resolution, but our data are consistent with published values. }
{These observations illustrate the cometary science that will be performed with the future ALMA interferometer.}


\titlerunning{\h2s, SO and CS interferometric imaging in comet Hale-Bopp}
\authorrunning{Boissier et al.} 
\maketitle

\section{Introduction}

Spectroscopic investigations of comets in the microwave wavelengths domain began in the 80's when sensitive instrumentation  was made available.
 This wavelength range proved to be a powerful tool to identify and measure the relative abundances of parent molecules released from comet nuclei from the observations of their rotational lines   (see, e.g., the review of  Bockel\'ee-Morvan et al. \citeyear{dbm2004}). 
However, little information was obtained on their spatial distribution in the  inner coma  because of the poor angular resolution (typically $10 \arcsec$ at most) offered by single dish (diffraction-limited) measurements.    
The study of the spatial distribution of parent molcules is of strong interest as inferences concerning the  outgassing processes at the nucleus surface can be  made. 
In addition, the study of the brightness radial distribution of molecular species can provide information on their origin, i.e. indicate whether they are indeed released by the nucleus or produced by some source in the coma. 
Such investigation has been performed for decades on radicals and atoms observed at visible or UV wavelengths (see the review of \citealt{feld2004}) and provided measurements of their photodissociative lifetime   and the lifetime of their parents. 
Spatial 1-D  mapping at high angular resolution has also been undertaken for a few molecules   (CO, HCN, H$_2$O) observed at infrared wavelengths in comets C/1996 B2 (Hyakutake) and/or C/1995 O1 (Hale-Bopp)  (e.g., \citealt{disanti01}; \citealt{mag99}).   
  
Mapping of rotational emission lines at high angular resolution was only attempted in a few comets, mostly  in comets Hyakutake and Hale-Bopp. Because of their exceptional brightness, interferometric imaging  of molecular lines  could be successfully performed (e.g., \citealt{wright}; \citealt{blake}; \citealt{woodney}). 
 We present here observations of three sulfur bearing species undertaken in comet Hale-Bopp in March 1997 using the IRAM Plateau de Bure interferometer at angular resolution of 1.6 to 3.5$\arcsec$.
 H$_2$S and SO have  so far only  been detected in comets in the millimetre or submillimetre domains \citep{dbm2004}. 
Our observations provide the first study of their spatial distribution. CS has been detected at high angular resolution   in the UV in several comets (\citealt{feld2004}).
 Its spatial distribution was studied in the microwave domain in comet Hale-Bopp using the Berkeley Illinois Maryland Association (BIMA) array (\citealt{sny01}).
 Our observations of two CS rotational lines usefully complement these studies. 
    
Section~\ref{obs} presents the data set. Interferometric images, spectral maps and autocorrelation (i.e., single-dish) spectra of H$_2$S, SO and CS lines 
are analysed and compared to investigate jet-like structures and temporal variability linked to nucleus rotation. 
In Sect.~\ref{sect:astro},  we show that a well-determined gravitational solution for comet Hale-Bopp orbit can be deduced from the astrometry provided by the Plateau de Bure continuum observations. 
Section~\ref{results}  focuses on the study of the radial distribution of the molecules, the model being described in Sect.~\ref{model}. A summary follows in Sect.~\ref{summary}.

\section{Observations}
\label{obs}

\subsection{Description}

\begin{table*}
      \caption[]{Log of the  observations and main characteristics.}
            \label{log}
	    \centering
         \begin{tabular}{l l  c c c c c c r c}
                 \hline\hline
            \noalign{\smallskip}
             Line & Date & UT & $r_h$ & $\Delta$ & Mode & $S/{T_{A}}^{~*\mathrm{a}}$  &Beam & $\textrm{Line area}^{\mathrm{b}}$ &
            $\Delta \textrm{v}^{\mathrm{c}}$ \\
             $\nu$ (GHz) & (1997) &  & (AU) & (AU) & & (Jy/K) & ($\arcsec$) &
            $\textrm{Jy km s}^{-1}$  &
            $\textrm{km s}^{-1}$ \\

            \noalign{\smallskip}
            \hline
            \noalign{\smallskip}
             CS $J$(2--1)   & March 12 & 8h47--15h30 & 0.98 & 1.36 & cross & 23.5  & 4.39 $\times$ 2.82 & 1.47 $\pm$ 0.09 &  --0.19 $\pm$ 0.02\phantom{0}    \\ 
              97.9809533   & & 8h41--14h03 & & & auto & 21.1  & 49.0 & 15.2 $\pm$ 0.2\phantom{0} & --0.06 $\pm$ 0.02\phantom{0} \\
             \hline
      \noalign{\smallskip}
             CS $J$(5--4) & March 12 & 6h31--15h30 &  0.98 & 1.36 & cross & 63.1 & 1.92 $\times$ 1.28 & 5.4 $\pm$ 0.8\phantom{0} & --0.10 $\pm$ 0.02\phantom{0}      \\ 
             244.9355565   & & 6h24-14h03 & & & auto & 30.4 & 19.7 & 213.2 $\pm$ 0.6\phantom{0} & --0.079 $\pm$ 0.003 \\
             \hline
      \noalign{\smallskip}
             \h2s{} $2_{20}$--$2_{11}$   & March 13 & 4h25--15h27 &  0.97 & 1.35 & cross & 37.6 & 1.99 $\times$ 1.53 & 3.51 $\pm$ 0.15 &   --0.05 $\pm$ 0.01\phantom{0}   \\ 
               216.7104350  & & 4h19-14h37 & &             & auto & 27.0 & 22.15 & 18.9 $\pm$ 0.3\phantom{0} & --0.06 $\pm$ 0.02\phantom{0} \\
             \hline
      \noalign{\smallskip}
             SO $N_{J}$(5$_{6}$--4$_{5}$)   & March 13  & 4h25--15h27 &  0.97 & 1.35 & cross & 37.6 & 1.93   $\times$ 1.5 & 0.44 $\pm$ 0.04 & --0.05 $\pm$ 0.09\phantom{0}     \\
              219.9494420  & & 4h19--14h37 & & & auto & 27.0 & 21.9 & 16.9 $\pm$ 0.3\phantom{0} & --0.13 $\pm$ 0.02\phantom{0} \\

            \noalign{\smallskip}
            \hline
         \end{tabular} 
\begin{list}{}{}
\item[$^{\mathrm{a}}$] Conversion factor averaged over the six antennas. Forward and beam 
efficiencies
are ($F_{eff}$, $B_{eff}$) = (0.93, 0.83), (0.89, 0.62), (0.89, 0.55) at 98, 220 and 245 GHz, respectively. 
\item[$^{\mathrm{b}}$] Intensity integrated over velocity measured in \oo~ spectra (autocorrelation mode),  and in the central pixel of the interferometric maps  (cross-correlation mode). 
The given  uncertainties do not include uncertainties in the absolute  calibration  ($\sim$ 15\% in both modes). 
\item[$^{\mathrm{c}}$] Line velocity shift defined as $\Delta v = \frac{ \sum_{i}T_{i}v_{i} } { \sum_{i}T_{i} }$, where $T_{i}$ and $v_{i}$ are the intensity and velocity of channel $i$, respectively
\end{list}
\end{table*}

Comet C/1995 O1 (Hale-Bopp) was observed in 1997 at the IRAM Plateau de Bure Interferometer situated in the french Alps \citep{gui+92}.
The observing campaign lasted 2 weeks between March 6 and March 22, at the time of the closest approach of the comet to the Earth (perigee was on March 22).  
The programme included interferometric mapping and single dish  measurements  in spectroscopic mode for several molecular species (CO, HCN, HNC, CS, \h2s, SO$_{2}$, SO, CH$_{3}$OH, H$_{2}$CO; \citealt{wink99}; \citealt{dbm2000}; \citealt{hen02}; \citealt{hen03}; \citealt{dbm2005}),  as well as observations of the dust and nucleus continuum emission  (\citealt{altenhoff}).

At the IRAM Plateau de Bure interferometer, observations with 1.3 and 3 mm receivers were conducted at the same time.    
The CS $J$(2--1) and $J$(5--4) lines (at 98.0 and 244.9 GHz, respectively) were observed in parallel on March 12. 
On March 13, the $2_{20}$--$2_{11}$ line of \h2s{} (at 216.7 GHz) and $N_{J}$(5$_{6}$--4$_{5}$) line of SO (at 219.9 GHz) were observed simultaneously in the same backend, taking benefit of double-sideband receivers. 
A log of the observations is presented in Table~\ref{log}. 
At this time, the instrument counted five 15-m antennas set in C1 compact  configuration so that baseline lengths projected onto the plane of the sky were between 22 m and 147 m. 
The comet was at a geocentric distance  $\Delta = 1.35-1.36$ AU and heliocentric distance $r_{h} = 0.97-0.98$ AU. 

The comet was tracked using orbital elements computed by D.K. Yeomans (JPL ephemeris solution 55) and the ephemeris program of the Institut de m\'ecanique c\'eleste et de calcul des \'eph\'em\'erides  (IMCCE, Observatoire de Paris) that takes into account planetary perturbations. 
First interferometric maps obtained on March 9 showed  that the intensities (of both continuum and line emissions) peaked  5--6$\arcsec$ northward with respect to the expected position.
The ephemeris was corrected by $+ 6 \arcsec$ in declination (Dec) for the observations performed on March 12 and 13. It turned out that the position  of the brightness peaks (continuum and lines) was offset by less than $\sim$2$\arcsec$  with respect to the pointed position. 
Single dish measurements are not affected by this pointing offset given the size of the primary beam at the observed frequencies (Half Power Beam Width, HPBW, of~22 to 50$\arcsec$, Table~\ref{log}).

The observing cycle was typically: pointing, focussing, 4 min of cross-correlation on the calibrators, 2 min of autocorrelation (\oo{} measurements, 1 min on source)  and 51 scans (1 min each) of cross-correlation on the comet. 
At the end of some cycles, additional autocorrelation spectra were obtained.  
Hereafter, autocorrelation observations will be refered to as ``\oo''  observations. 
The OFF position in ON--OFF (position switching)  observations was at 5$\arcmin$ from the ON position. 

In 1997, the correlator consisted of 6 independent units, adjustable in resolution and bandwidth. On March 12, two units were used for observing CS, the other ones were used for CH$_3$OH lines near 241 GHz. 
On March 13, four units were used for measuring the continuum emission (\citealt{altenhoff}), and one unit covered  the $J$(1--0) HNC line observed with the 3 mm receivers (\citealt{dbm2005}). 
Another one covered both the   $2_{20}$--$2_{11}$ \h2s{} and $N_{J}$(5$_{6}$--4$_{5}$) SO   lines present in lower and upper receiver sidebands. 
The units used for CS, \h2s{} and SO lines were divided into 256 channels with 78 kHz separation, providing an effective velocity resolution of $\sim 0.31  \textrm{~~km~s}^{-1}$ for CS $J$(2--1), $\sim 0.12 \textrm{~~km~s}^{-1}$ for CS $J$(5--4) and $\sim 0.13  \textrm{~~km~s}^{-1}$ for \h2s and SO lines (the resolution is 1.3 times the channel spacing).

The frequency calibration was checked by observing the  reference source W3OH. Interferometric observations were calibrated using the standard IRAM software CLIC.
The amplitude and phase calibrator was BL Lac (2200+420) which flux density was determined observing MWC349. 
In \oo~mode, the intensity scale is the antenna temperature ${T_{A}}^{*}$.
The conversion factor from ${T_{A}}^{*}$ to flux density and the HPBW  of the primary beam at the frequencies of CS, \h2s and SO lines are given in Table~\ref{log}. 
The uncertainty in flux calibration is about 15\%, both in interferometric and \oo{} mode.

\subsection{\oo{} spectra}
\label{onoff}
\begin{figure}
\centering
\includegraphics[angle=-90,scale=0.8]{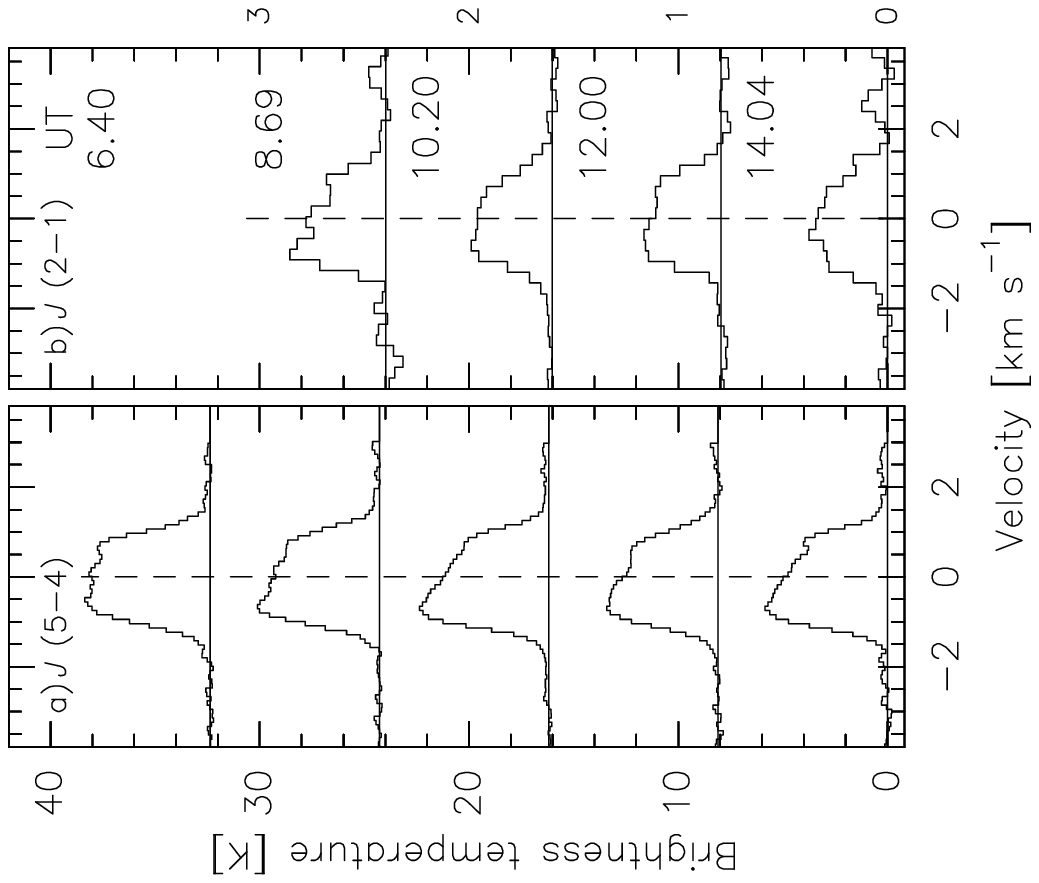}
\\[0.5cm]
\includegraphics[angle=-90,scale=0.7]{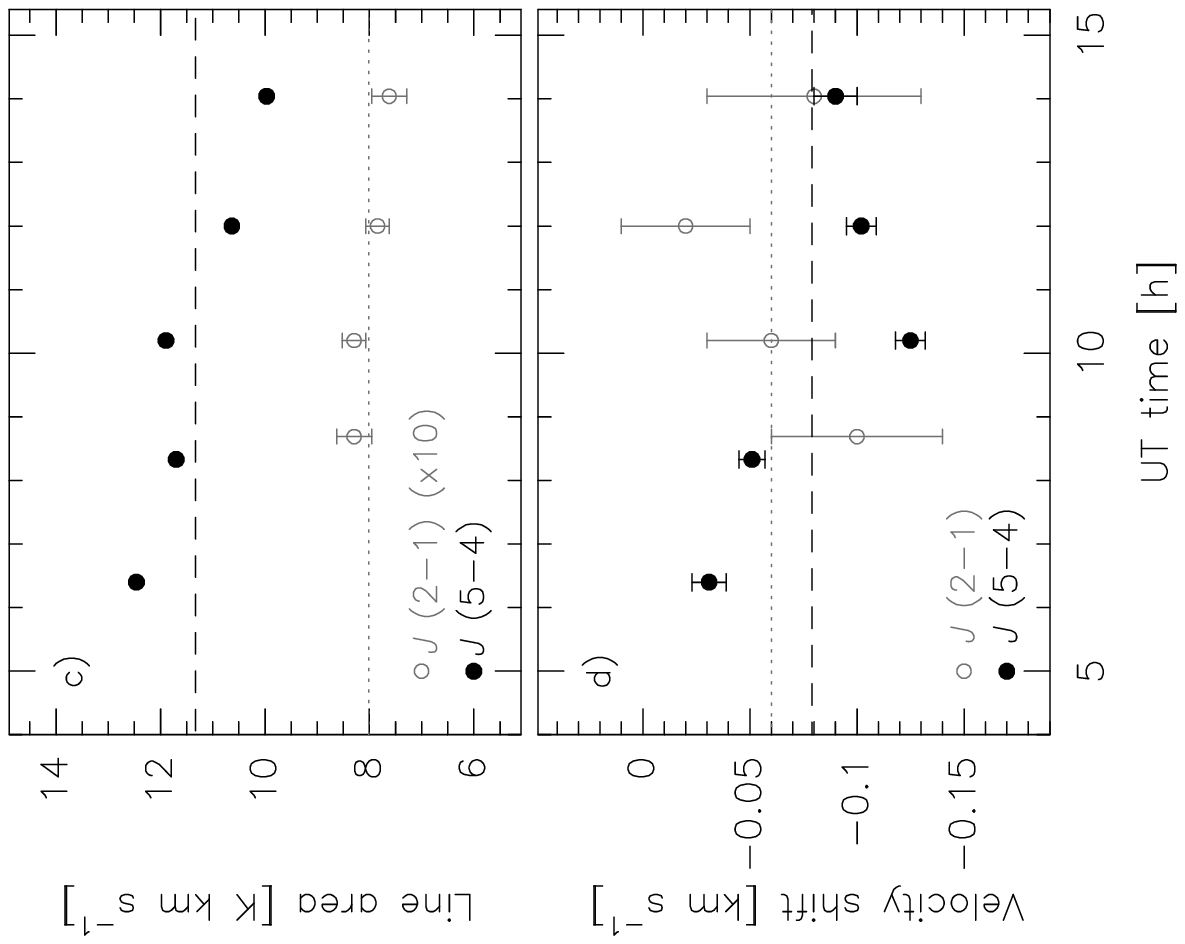}
\caption{\textbf{\emph{a--b}}: \oo~spectra of CS $J$(5--4) (244.9 GHz)  (\emph{a}) and CS $J$(2--1)  (98.0 GHz) (\emph{b}) lines recorded on 
  March 12, 1997. The spectra, in brightness temperature scale, are shifted 
  vertically according to observation UT 
  time (given in the top right corners).  Horizontal lines correspond to zero
  intensity. The integration time is 2 min (CS $J$(5--4) at 6.40 h and 14.04 h and CS  $J$(2--1) at 8.69 h and 14.04 h) or 4 min.  The velocity scale is with respect to the comet rest velocity. 
 \textbf{\emph{c}}:  Line area as  a fonction of UT time. Black and grey symbols are for $J$(5--4) and  $J$(2--1) lines, respectively. The black dashed (respectively grey dotted) line shows mean value  for $J$(5--4)~(respectively $J$(2--1)) line.
  \textbf{\emph{d}}:  Line velocity shift as 
  a fonction of UT time. Black and grey symbols are for $J$(5--4) and  $J$(2--1) lines, respectively. The black dashed (respectively grey dotted) line shows mean $\Delta v$ value for $J$(5--4)~(respectively $J$(2--1)) line.}
\label{evolCS}
\end{figure}

\begin{figure}
\centering
\includegraphics[angle=-90,scale=0.8]{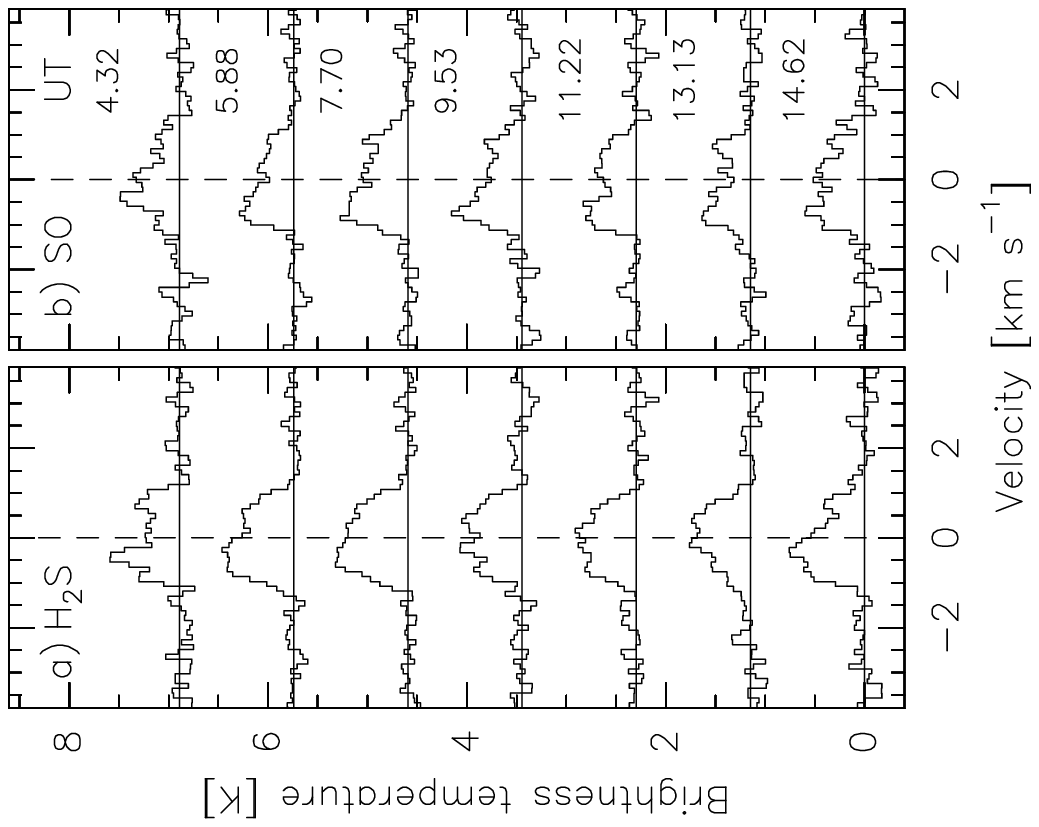}
\\[0.5cm]
\includegraphics[angle=-90,scale=0.7]{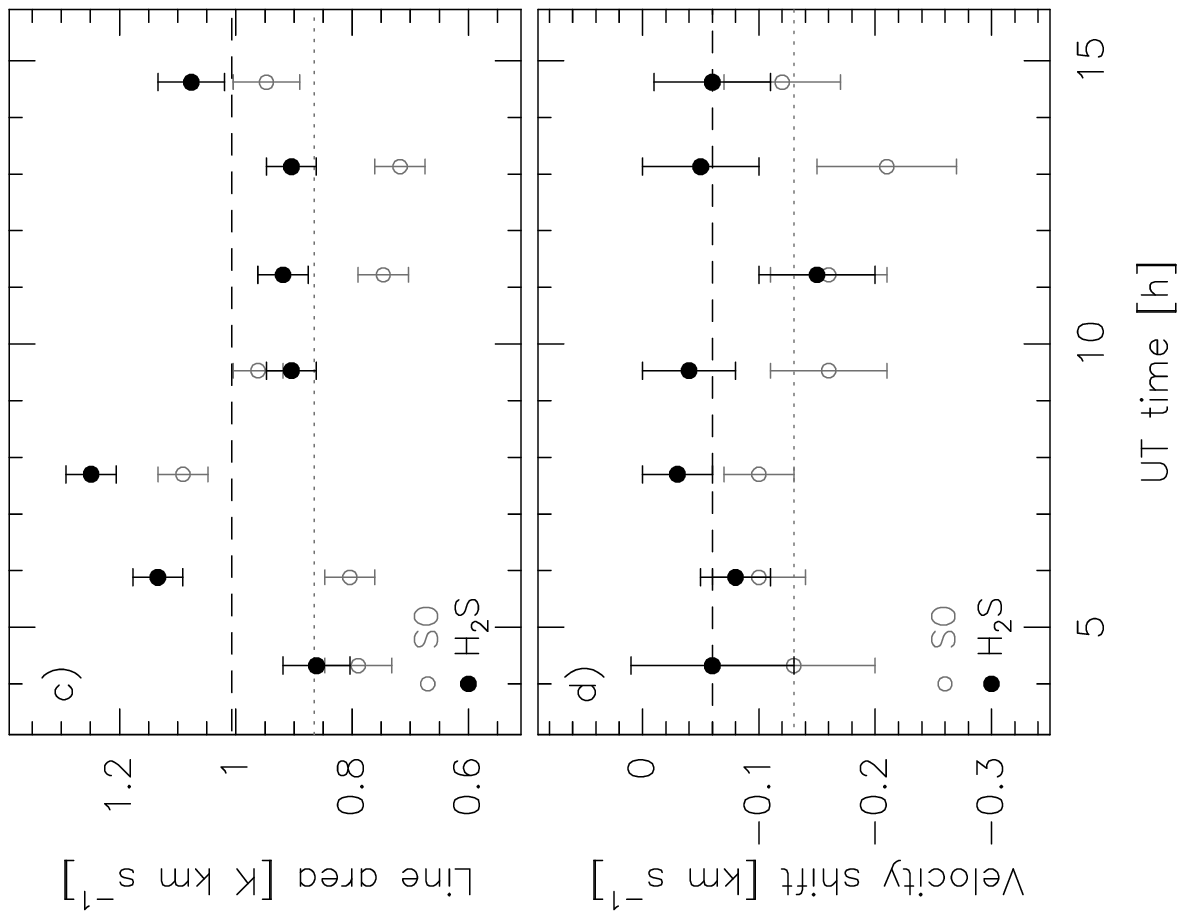}
\caption{\textbf{\emph{a--b}}: \oo~spectra of   \h2s{} $2_{20}-2_{11}$ (216.7 GHz)  (\emph{a}) and SO    $N_{J}$(5$_{6}-4_{5}$) (219.9 GHz) (\emph{b}) lines recorded on   March 13, 1997.
 The spectra, in brightness temperature scale, are shifted   vertically according to observation UT  time (given in the top right corners).  
Horizontal lines correspond to zero  intensity. 
The integration time is 4 min   for all spectra, but those at 4.32 h and 14.62 h UT for which it is 2  min.
 The velocity scale is with respect to the comet rest velocity.  
  \textbf{\emph{c}}:  Line area as   a fonction of UT time. Black and grey symbols are for \h2s{} and SO,  respectively.
 The black dashed  (respectively grey dotted) line shows mean value   for \h2s~(respectively SO) line.
 \textbf{\emph{d}}:  Line velocity shift as a fonction of UT time. Black and grey symbols are for \h2s{} and SO,  respectively.
 The black dashed   (respectively grey dotted) line shows mean $\Delta v$ value for \h2s~(respectively SO) line.}
\label{evol}
\end{figure}

The \oo{} spectra obtained independently by the five antennas were co-added. 
All lines were detected with a signal-to-noise ratio $>$ 10 in single scans, so that the evolution of the  line characteristics along the observation period could be studied to investigate possible changes linked to nucleus rotation.
Since the  rotation period of the nucleus of comet Hale-Bopp is about  11.35 h (e.g., \citealt{jorda99}), the nine \h2s and SO spectra obtained between 4h19 and 14h37 UT (Table \ref{log}) spread over almost a full rotation period. 
Similarly, the observations of CS $J$(5--4) cover 80\% of the nucleus period.
 Technical problems prevented us from obtaining as much time coverage for CS $J$(2--1). 
Spectra as a function of time are shown in Fig.~~\ref{evolCS}--\ref{evol}. 

 As shown in Figs. 1--2, no significant temporal changes that could be related to nucleus rotation are observed for the line areas.
We also plot in Figs. 1--2 the temporal evolution of the line centroids in nucleus velocity frame, hereafter referred to as line velocity shifts or velocity offsets (see footnote (c) of Table 1).
For all molecules, velocity offsets  are always negative, taking into account error bars. 
Given the phase angle on March 12--13 ($\sim 45^{\circ}$), this is consistent with a Sun/anti-Sun asymmetry in the distribution of the molecules, with more molecules towards the Sun.
The velocity offsets of the H$_{2}$S ($2_{20}$--$2_{11}$), SO $N_{J} (5_{6}$--$4_{5})$ and CS $J$(2--1) lines present some fluctuations but no clear evidence for a 11.35~h periodic modulation (Figs.~\ref{evolCS}--~\ref{evol}). 
In contrast, the velocity offset of the CS $J$(5--4) line, measured with much better accuracy (less than 0.01 km s$^{-1}$ rms), presents a clear trend of variation with rotation phase, with a minimum on  March 12 10h15 UT at $-$0.125 km s$^{-1}$, and maximum  at $-$0.03 km s$^{-1}$ observed 4 h earlier. 

These measurements have to be compared to the strong modulation ($\sim$ 0.3 km s$^{-1}$ total amplitude) observed on March 11 for the CO $J$(2--1) and $J$(1--0) velocity offsets (the lowest velocity offset of the CO  $J$(2--1) line was near 8h30 UT).
 This modulation has been interpreted as caused by a rotating CO jet emanating from a source near the nucleus equator that was active day and night (\citealt{hen02}; \citealt{hen03}; Bockel\'ee-Morvan et al. {\it in preparation}; \citealt{Boi06}), as in mid-March the  nucleus rotation axis was near the plane of the sky. 
The velocity offset of the CO lines took negative to positive values because the velocity vector of the CO-jet with respect to Earth rotated during the course of the observations.
Comparing CS and CO lines evolution, one can conclude that their evolution are not in phase, and thus that the CO and CS sources might be different.

\begin{figure}
\resizebox{\hsize}{!}{\includegraphics[angle=-90]{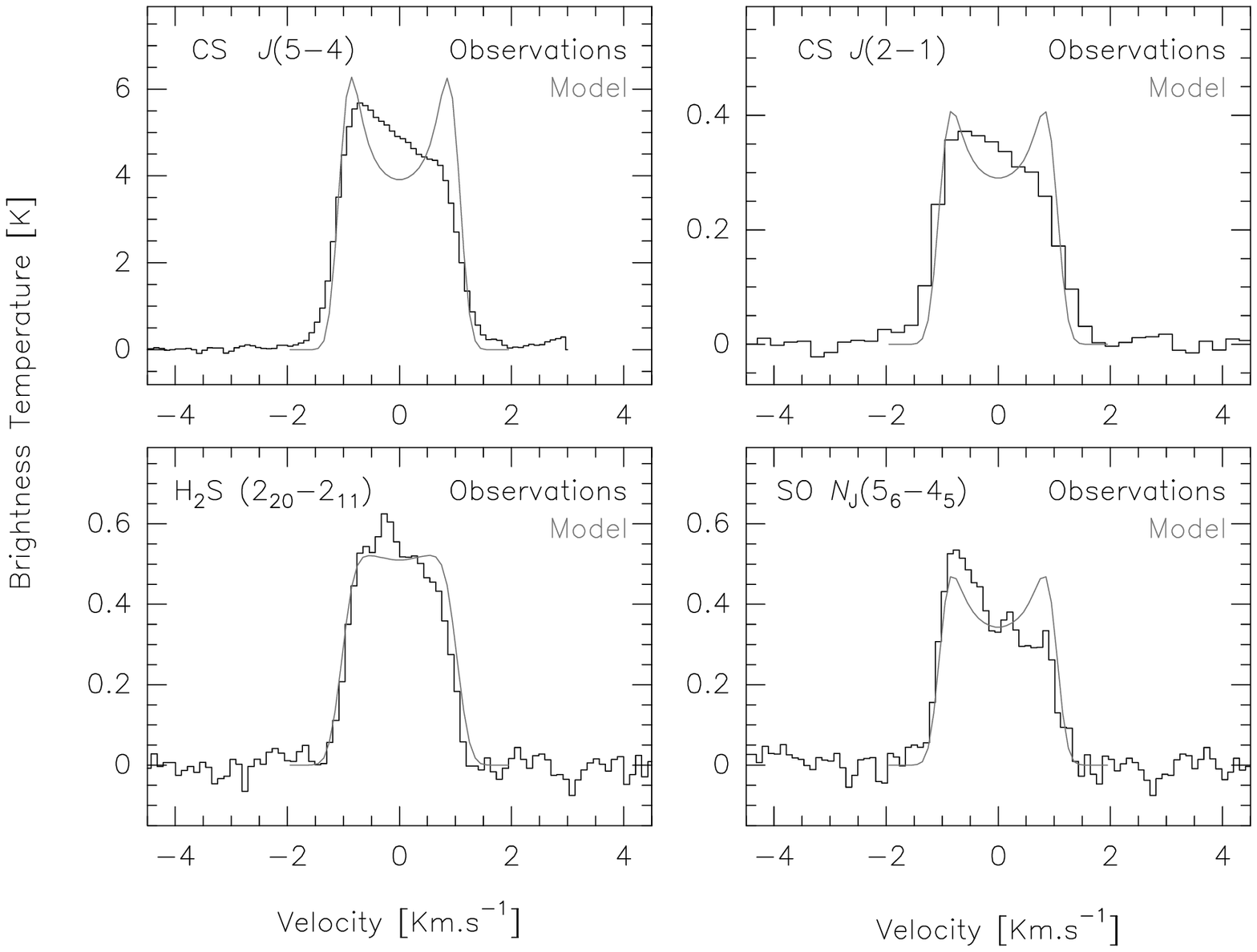}}
\caption{CS $J$(5--4) (244.9 GHz, the integration time (ON+OFF) is 16 min), $J$(2--1) (98 GHz, 12 min), \h2s{} ($2_{20}$--$2_{11}$) (216.7 GHz, 24 min) and SO $N_{J}$(5$_{6}$--$4_{5}$) (219.9 GHz, 24 min)  ON--OFF spectra obtained on March 12 and 13, averaging all scans.  
A typical synthetic profile resulting from the model described in Sect.~\ref{model}  is overplotted in grey line.}
\label{spetot}
\end{figure}

The average spectra obtained on March 12 and 13 are shown in Fig.~\ref{spetot}.
Their characteristics are summarized in Table \ref{log}. 
Comparing the shapes of the lines and their Doppler shift, we can notice that the SO line is more asymmetric than the \h2s{} line. 
This will be further discussed in Sect.~\ref{interf}.  

\subsection{Interferometric observations}
\label{interf}
\subsubsection{Principle of interferometric imaging}
  
An interferometer measures  the 2D Fourier Transform (FT) of the brigthness distribution on the sky. At a time $t$ a  pair of antennas provides a point of the FT (a complex number called \emph{visibility}) at the coordinates ($u,v$) in the Fourier Plane (also called \emph{uv--plane}), corresponding to the projection on the sky of the baseline (separation between antennas).
 The baseline projection evolves due to Earth rotation so that a single  pair  of antennas probes several points of the FT during the observations. 
At the end of the observing run, the visibilities are stored in a  $uv$-table.
 The final observed sample in the Fourier plane is called \emph{uv--coverage}.

Interferometric maps are obtained by computing the inverse Fourier Transform of the measured signal. 
However, because in practice it is impossible to probe the whole Fourier plane, interpolation is required.
 The asymmetry in the \emph{uv--coverage} results in an elliptic  clean beam.
The clean beam (hereafter referred to as interferometric beam) is the central part of the FT of the $uv$--coverage.
Direct analysis of the visibilities allows us to avoid the uncertainties due to the interpolation.

\subsubsection{CS, \h2s{} and SO interferometric maps}
\label{intobs}

\begin{figure}
\resizebox{\hsize}{!}{\includegraphics[angle=-90]{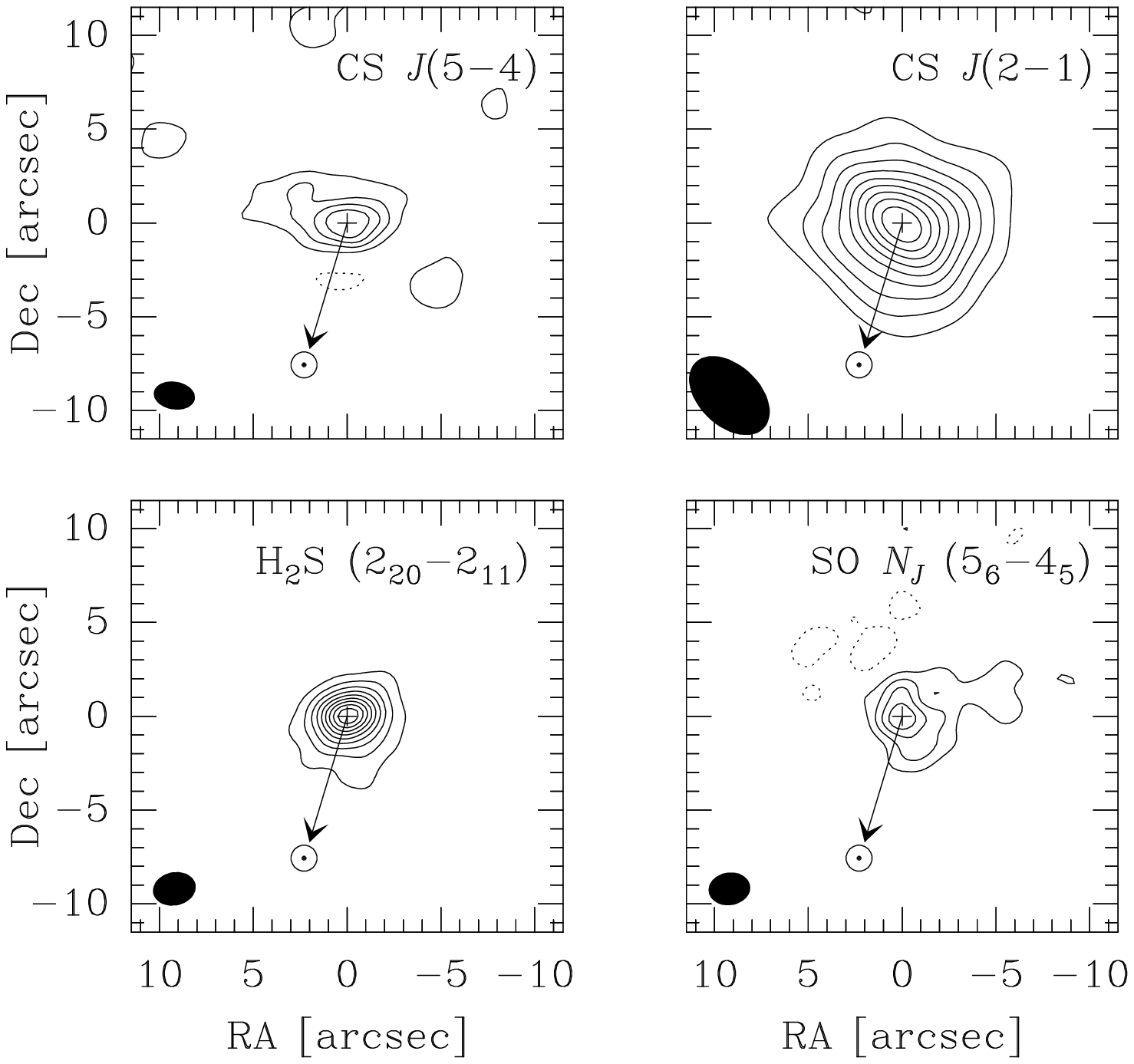}}
\caption{Interferometric maps of CS $J$(5--4) (244.9 GHz, 5.9 h integration time), $J$(2--1) (98 GHz, 4.7 h),  \h2s{} (216.7 GHz, 6.6 h) and SO (219.9 GHz, 6.6 h) lines on March 12 for CS and 13 for \h2s~and SO. Spectral channels covering a velocity interval of 3.67, 4.32, 2.8 and 3.1 $\textrm{~km~s}^{-1}$ centred on comet rest velocity have been considered for  CS $J$(5--4), $J$(2--1), \h2s{} and SO, respectively. 
The interferometric beam is shown in the left bottom corner (see values in Table~\ref{log}). 
The  Sun direction is indicated  by the arrow. 
Contours correspond to 10\% (respectively 20\%) of the peak  intensity measured on the centre of the maps for \h2s{} and CS $J$(2--1) (respectively SO and CS $J$(5--4)).}
\label{maps}
\end{figure}

 \begin{table*}
      \caption[]{CS, \h2s{} and SO maximum line brightness positions.}
         \label{position}
         \centering
         \begin{tabular}{ c c c c c c c}
           \hline \hline
            \noalign{\smallskip}
          Line&  Date  & UT$^{\mathrm{a}}$  & RA$^{\mathrm{b}}$ & Dec$^{\mathrm{b}}$ & (O-cont)$^{\mathrm{c}}$ & (O--C)$^{\mathrm{d}}$\\
           &  & &  & &  ($\delta$RA,$\delta$Dec) &  ($\delta$RA,$\delta$Dec) \\
           &  & h:min:s & h:min:s & $^{\circ}$:$\arcmin$:$\arcsec$ & ($\arcsec$,$\arcsec$) & ($\arcsec$,$\arcsec$)\\

            \noalign{\smallskip}
            \hline
            \noalign{\smallskip}
            CS $J$(2--1)& March 12 & 15:00:00 & 22:41:28.643 & 41:36:31.11 & $^{\mathrm{e}}$  &  (+0.42,+2.13)  \\
            CS $J$(5--4)& March 12 & \phantom{0}7:00:00 & 22:38:43.424 & 41:24:11.52 & $^{\mathrm{e}}$  &  (+0.37,+2.48)  \\
            \h2s& March 13 & \phantom{0}8:00:00 & 22:47:25.371 & 42:02:02.88 & (+0.06,$+$0.05) &  (+0.92,+2.87)  \\
            SO& March 13 & \phantom{0}5:00:00 & 22:46:21.868 & 41:57:35.54 & ($-$0.03,$-$1.34) &  (+0.83,+1.48)  \\
            \noalign{\smallskip}
            \hline
         \end{tabular}
\begin{list}{}{}
\item[$^{\mathrm{a}}$] UT time at which coordinates are given. 
\item[$^{\mathrm{b}}$] Apparent geocentric coordinates of the  brightness peak measured on interferometric maps (``O''). Gaussian fits in the $uv$--plane provide  similar absolute positions.
\item[$^{\mathrm{c}}$] Offset in RA and Dec between molecules brightness peak position (``O'') and 1.3-mm dust  continuum peak position (``cont'', \citealt{altenhoff}). Continuum observations were not performed on March 12.
\item[$^{\mathrm{d}}$] Offset in RA and Dec between molecules brightness peak position   (``O'') and the comet expected position (``C'')  based on astrometric measurements from 27 April 1993 to 4 August 2005 (JPL ephemeris solution 220).
\item[$^{\mathrm{e}}$] Continuum observations were not performed on March 12 (\citealt{altenhoff}).
\end{list}
\end{table*} 

\begin{figure}
\resizebox{\hsize}{!}{\includegraphics[angle=-90]{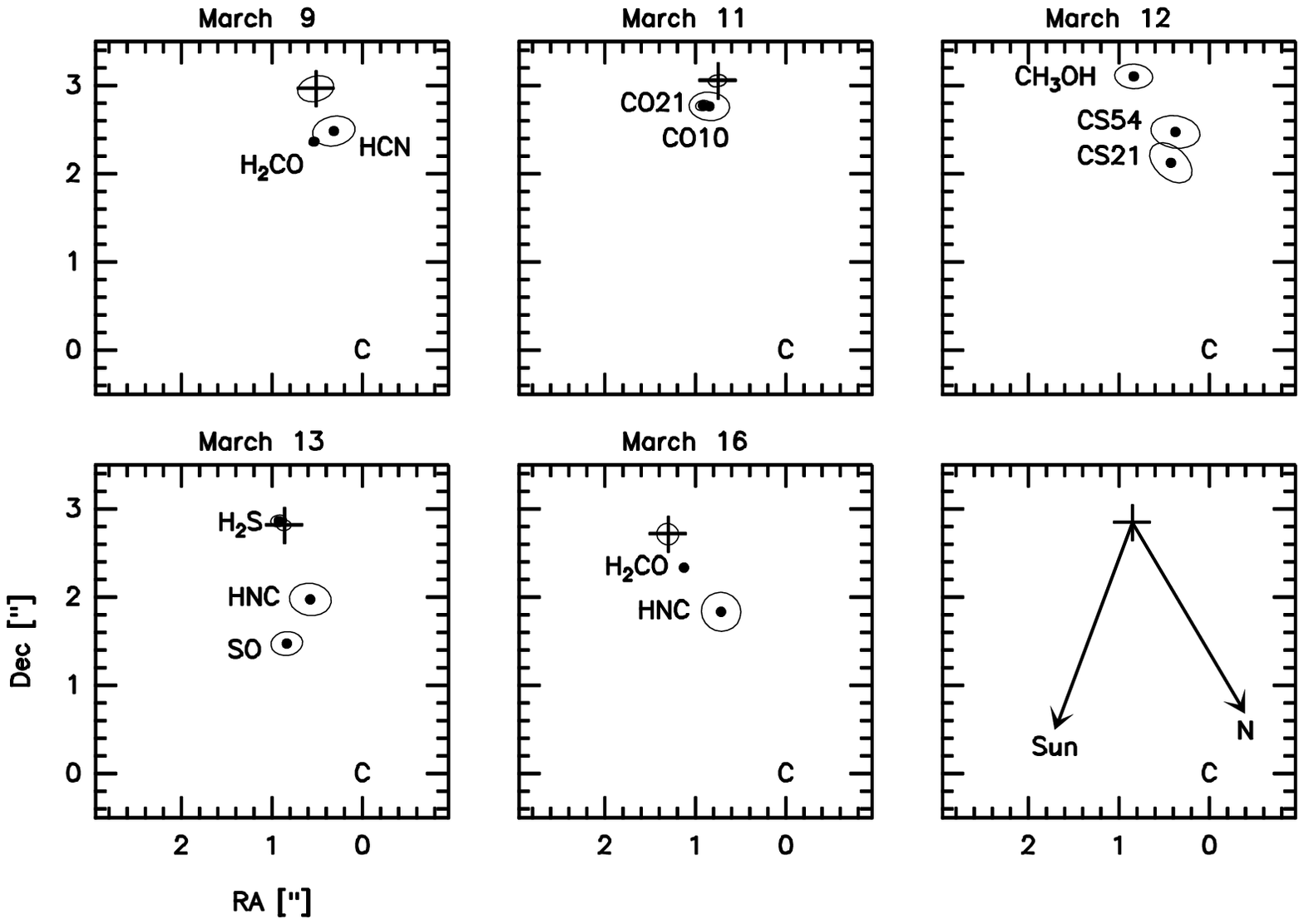}}
\caption{Position of maximum brightness peaks in molecular maps (plain circles) and 1.3-mm continuum maps (crosses, \citealt{altenhoff}) with respect to the ephemeris (C) from JPL solution 220.
 Surrounding ellipses correspond to  interferometric beam divided by the signal-to-noise ratio on peak intensity, and indicate the accuracy of the astrometric measurements.
The bottom-right panel shows the average continuum offset (cross) and the Sun and rotation axis (N is the nucleus North pole) directions projected onto the plane of the sky.}
\label{offsets}
\end{figure}

\begin{figure}
\resizebox{\hsize}{!}{\includegraphics[angle=-90]{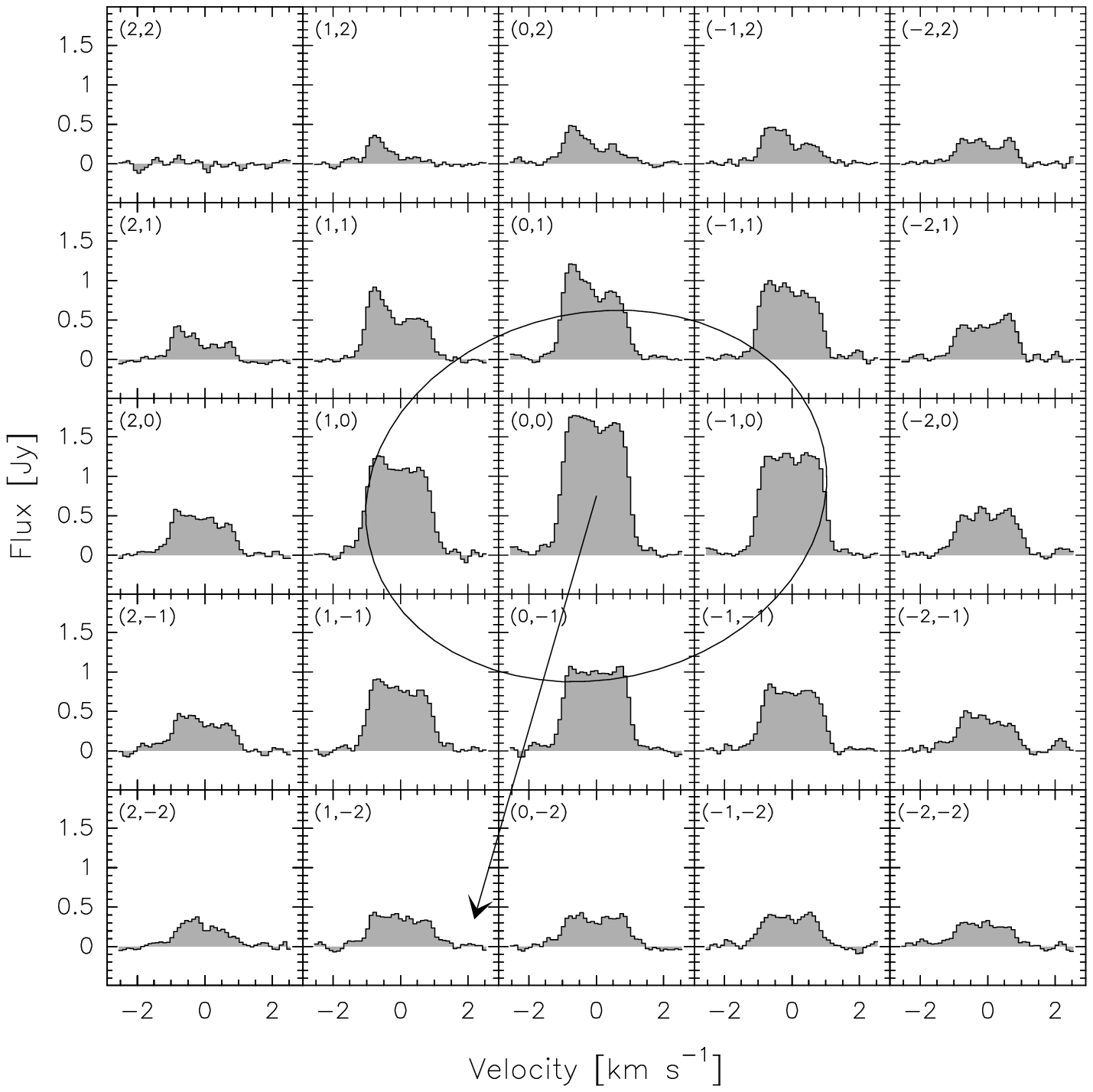}}
\caption{Spectral map of the H$_{2}$S ($2_{20}$--$2_{11}$) line. The integration time is 6.6 h.
 The position of the spectra (RA and Dec offsets in arcsec with respect to the position of maximum brightness) are given in the upper-left corner of the boxes.
The ellipse represents the interferometric beam and the arrow indicates the Sun direction projected on the plane of the sky. 
At northern positions, the blueshifted side of the line is stronger than the redshifted side. 
Anywhere else the line is overall symmetric.}
\label{smh2s}
\end{figure}

\begin{figure}
\resizebox{\hsize}{!}{\includegraphics[angle=-90]{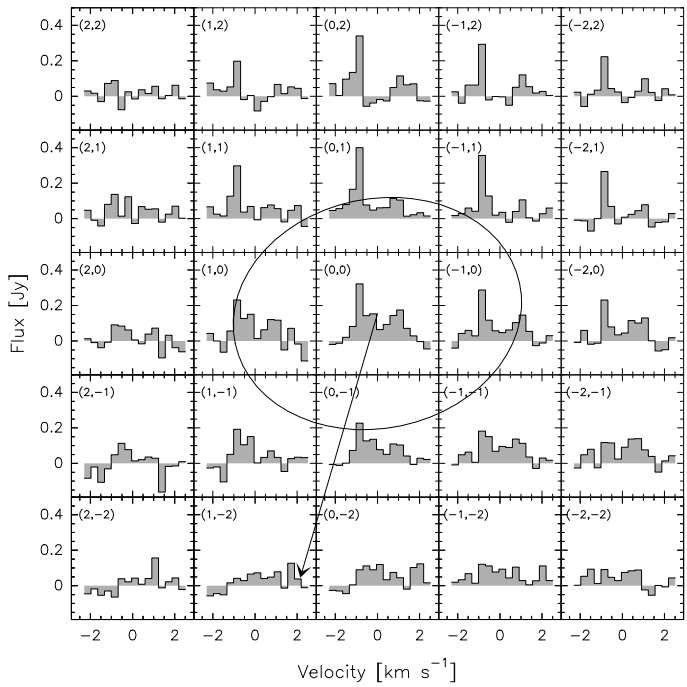}}
\caption{Same figure as Fig~\ref{smh2s} for  SO  $N_{J}$(5$_{6}$--4$_{5}$) line. 
The integration time is 6.6 h. 
In order to improve the signal to noise ratio the spectral resolution has been degraded by a factor 3 and is here about 0.32~km~s$^{-1}$. 
The line profile is strongly asymmetric at northern positions, mainly at negative RA offsets.}
\label{smso}
\end{figure}

\begin{figure}
\resizebox{\hsize}{!}{\includegraphics[angle=-90]{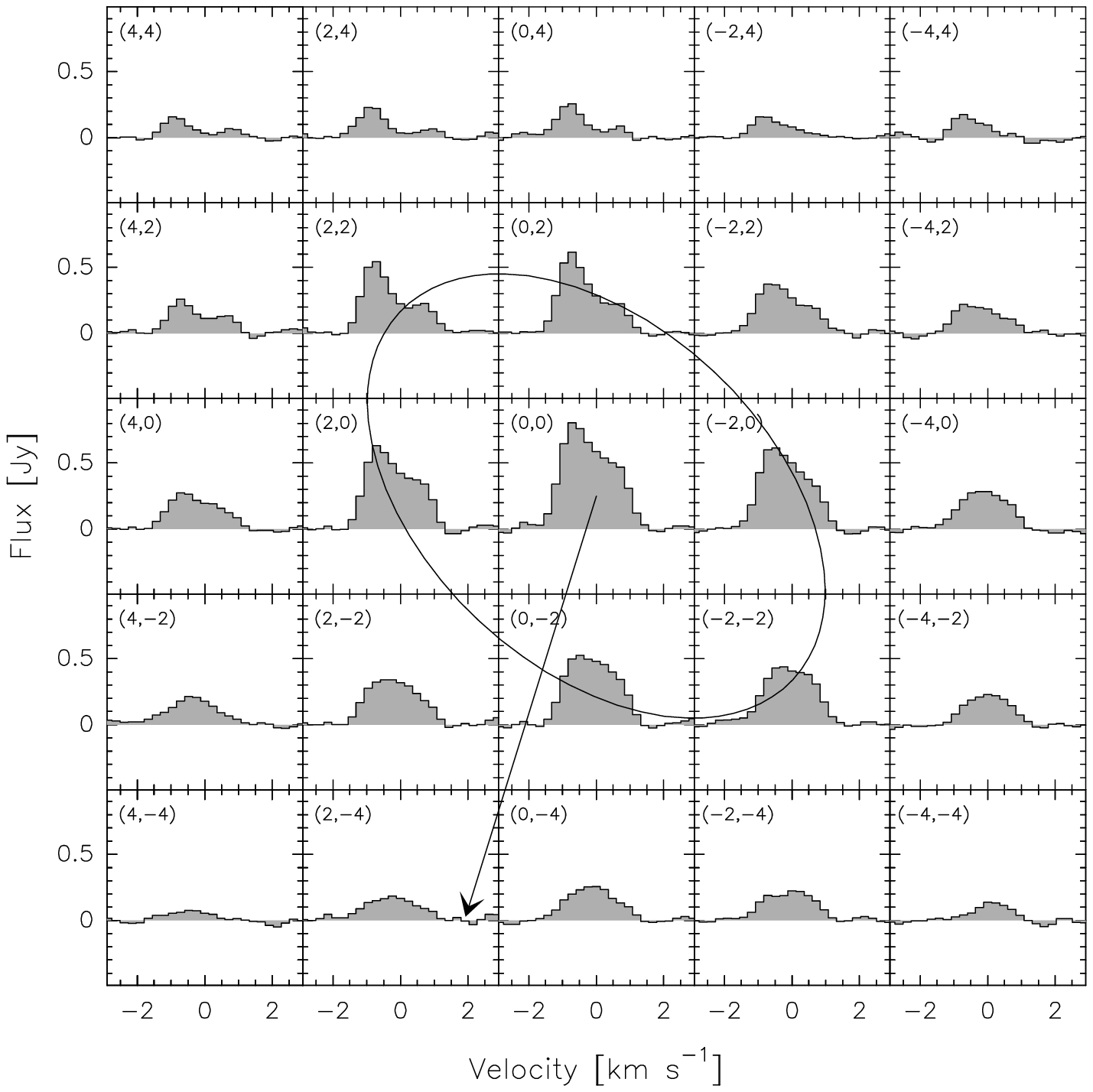}}
\\[0.3 cm]
\resizebox{\hsize}{!}{\includegraphics[angle=-90]{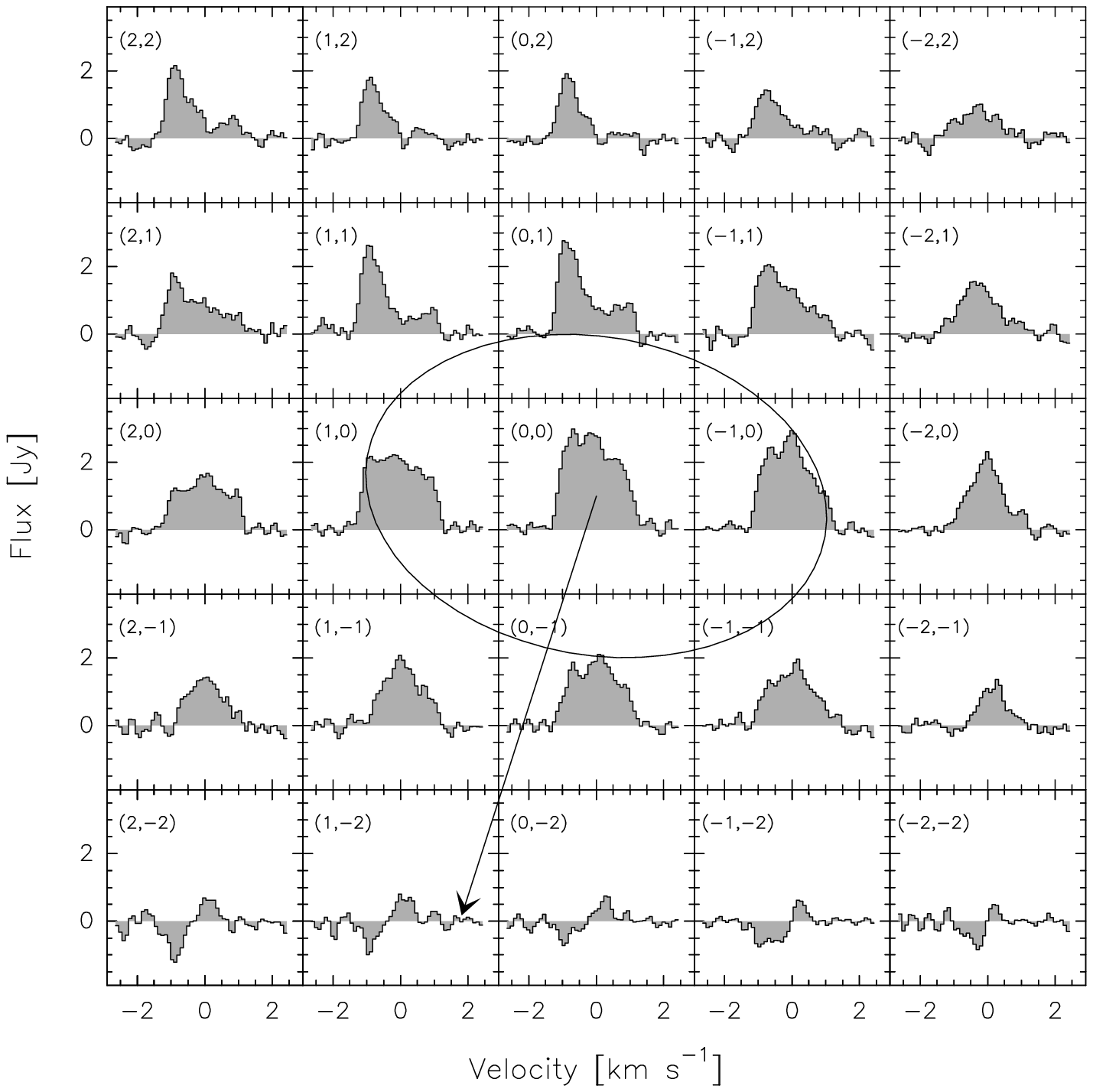}}
\caption{Same figure as Fig.~\ref{smh2s} for CS $J$(2--1) (top) and $J$(5--4) (bottom).
The integration time is 4.7 and 5.9 h, respectively. 
At positive Dec offsets (and especially at positive RA), the blueshifted side of  the lines is more intense; spectra are globally symmetric at negative Dec offsets.
In addition, at zero and negative Dec offsets, $J$(5--4) spectra present a narrow component peaking at $v \sim 0$. 
Note that the lack of flux in the  Dec=$-2 \arcsec$ row of CS $J$(5--4) spectral map  may be due to the lack of short spacings in the $uv$-plane.}
\label{sm54}
\end{figure}

The interferometric maps, obtained  by co-adding the velocity channels where line emission is present, are shown in Fig.~\ref{maps}. 
The original images have been shifted  so that the maximum brightness peaks at the centre of the maps.
 The main characteristics of these maps (intensity at maximum brightness, absolute and relative coordinates of the peak brightness (O) with respect to current ephemeris (C) and with respect to the position of maximum continuum emission (Cont) measured simultaneously) are summarized in Tables~\ref{log} and~\ref{position} and Fig.~\ref{offsets}. 
Spectral maps are shown in Figs.~\ref{smh2s}--\ref{sm54}. 
The following remarks can be done : 

\begin{itemize}
\item  The SO spatial distribution is more extended than  that of \h2s. 
The line flux in the   interferometric beam  is indeed eight times lower for SO (Table~\ref{log}), while SO and \h2s lines have comparable intensities in the primary beam,  as measured in autocorrelation mode (Fig.~\ref{spetot} and Table~\ref{log}). 
Radial distributions will be studied in detail in Sect.~\ref{results}.    
\item The points of maximum line brightness  do not coincide with the nucleus position provided by the ephemeris (C), found to be 1.5 to 3$\arcsec$ southward (Fig.~\ref{offsets}). 
Similar (O--C) offsets (by typically 1$\arcsec$ in RA and 3$\arcsec$ in Dec) are observed for  other lines (CO, HCN, HNC, H$_2$CO, CH$_3$OH) mapped at Plateau de Bure (\citealt{altenhoff}, Bockel\'ee-Morvan et al. {\it in preparation}). 
Continuum maps obtained simultaneously at Plateau de Bure also show a significant trend to peak northward of the ephemeris position, with (Cont--C) offsets  in Dec $\sim$+3$\arcsec$ (\citealt{altenhoff}). 
\item The positions of \h2s and CO maximum emission match the continuum position. No continuum data were obtained on March 12 at the date of CH$_3$OH observations, but almost identical (Cont--C) offsets were measured on March 9, 11 and 13 (Fig.~\ref{offsets}). 
Assuming that this (Cont--C) offset applies on March 12, the methanol emission peak also matches the continuum position. 
In contrast, the point of maximum brightness of other molecules  (namely HCN, H$_{2}$CO, CS, SO and HNC) lie at intermediate positions between the continuum and the ephemeris. 
\item  In the spectral maps (Figs.~\ref{smh2s}--\ref{sm54}),  the lines exhibit larger blueshifts at northward positions with respect to the centre of brightness in the maps. 
As already noted, blueshifted lines are expected for preferential outgassing towards the Sun. 
Taking into account the Earth-Comet-Sun geometry, stronger blueshifts towards northern positions is also consistent with a day/night asymmetry. 
Indeed, though the Sun is towards South in the maps (PA $\sim 160^{\circ}$), a larger number of molecules released from the nucleus illuminated hemisphere are observed to outflow northward rather than southward because the phase angle is $\sim 45^{\circ}$. 
This was checked by synthesizing spectral maps using the model detailed in Sect.~\ref{model} and a Sun/anti-Sun aymmetry in the distribution of the molecules in the coma. 
  \item There are however significant differences between \h2s and SO and CS spectral maps. 
 The \h2s spectra  at northern positions are less asymmetric than those of SO and CS at the same positions. 
 In addition, the CS $J$(5--4) spectra at centre and southern positions exhibit excess emission at Doppler velocities around 0 km s$^{-1}$. 
At these velocities, the emitting molecules are moving in (or near) the plane of the sky.  
\end{itemize}

The molecules can be then grouped in two categories: those whose brightness centres coincide with the continuum (\h2s, CO, CH$_3$OH) and whose spectral maps (averages over about one rotation period) show mainly evidence for a global day/night asymmetry.
Molecules of the second group (CS, SO, but also HCN, HNC and H$_2$CO) have their brightness centres offset towards South with respect to the continuum and may present, in addition to a diurnal asymmetry, a jet-like structure outflowing towards South in the plane of the sky, as observed for CS on its $J$(5--4) spectral maps. 
Comet Hale-Bopp exhibited in March 1997 a strong dust jet pointing southwest in visible images (\citealt{jorda99}) and interpreted as originating from high latitude regions on the nucleus surface (the rotation axis was at PA $\sim 210^{\circ}$ and aspect angle $\sim 70^{\circ}$ in mid-March 1997, i.e. almost in the plane of the sky, \citealt{jorda99}). 
We suggest that molecules of the second group present a gaseous counterpart of this dust jet. 

\begin{figure}
\resizebox{\hsize}{!}{\includegraphics[angle=-90]{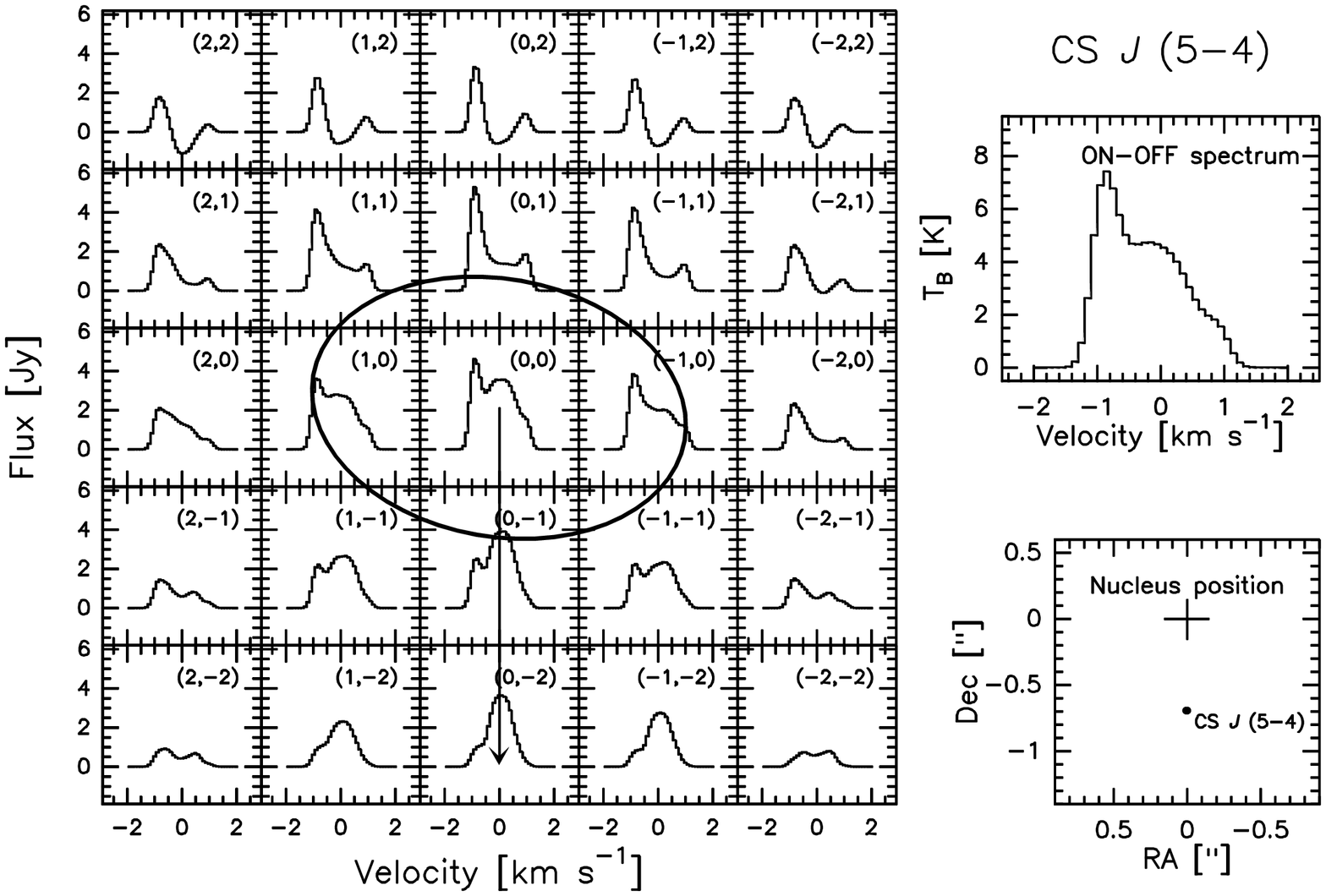}}
\\[0.3 cm]
\resizebox{\hsize}{!}{\includegraphics[angle=-90]{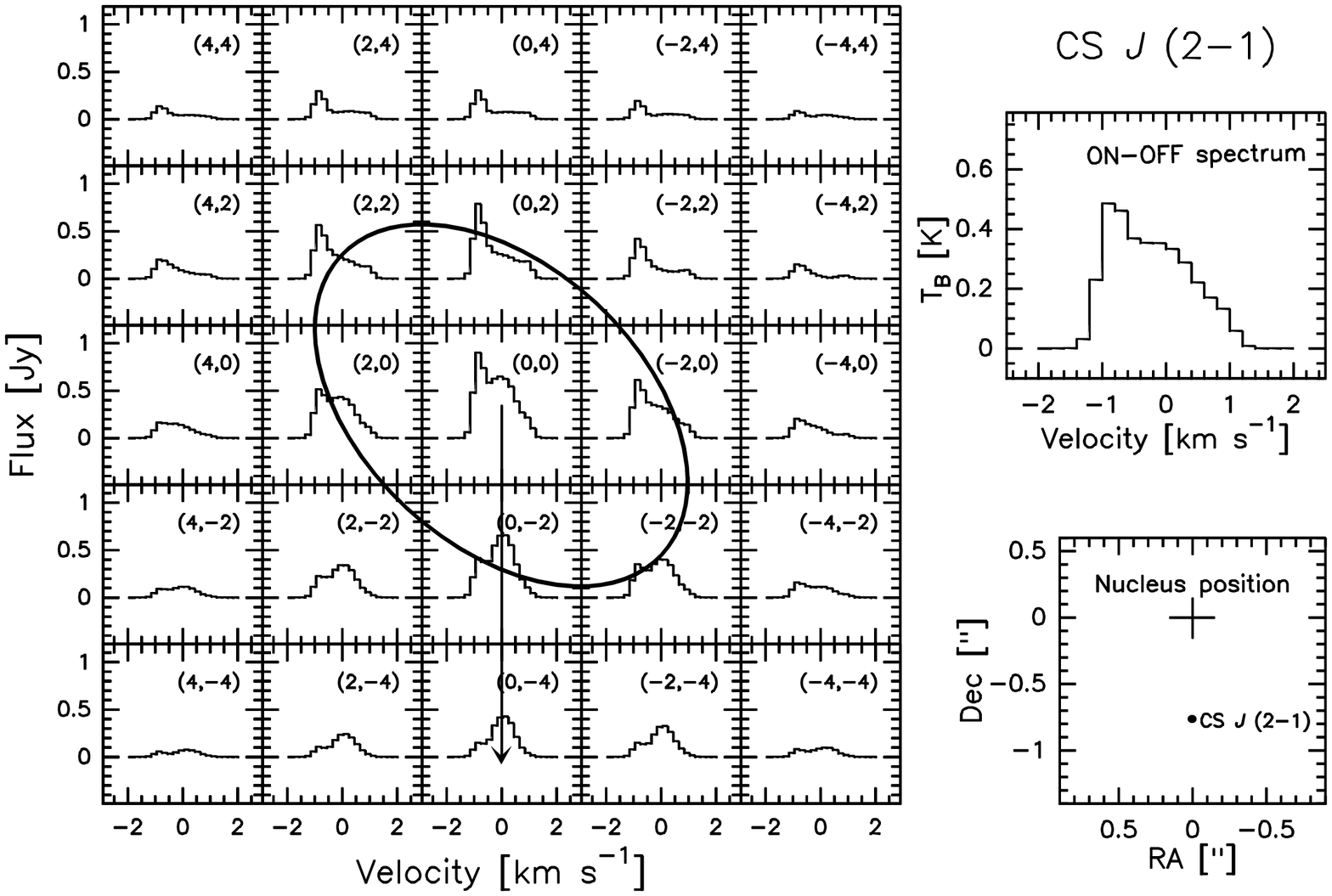}}
\caption{Synthetic spectral maps (left panel), \oo{} spectra (right upper panel) and positions of the brightness maximum with respect to the nucleus (right lower panel) for CS $J$(5--4) (top) and $J$(2--1) (bottom) lines as obtained with a model including a jet like structure southward  and a day/night asymmetry in the coma.
 The jet is 50$^{\circ}$ wide, and contains about 25 \% the total amount of CS. 
The CS production rate is 5  times stronger in the day side of the nucleus. 
The arrow indicates the projection of the Sun direction in the plane of the sky. 
The phase angle is 45$^{\circ}$.}
\label{allmodel}
\end{figure}

On the basis of synthesized maps obtained by describing the coma with a combination of conical structures of various opening angles and, for the continuum, a point-source mimicking nucleus thermal emission, we conclude that:
 1) the strong offset (Cont--C) $\sim$ 3$\arcsec$ in Dec cannot be explained by asymmetries in the dust coma, assuming a 50\% nucleus contribution to the thermal flux as  derived by \cite{altenhoff};
 the centre of brightness of the continuum maps should be very close (within 0.3$\arcsec$) to the actual position of the nucleus; 
2) Sun/anti-Sun asymmetries in the outgassing  that are compatible with observed blueshifts cannot displace the centre of molecular line brightnesses by more than a fraction of arcsec, thereby explaining why molecules of the first group have their centre of brightness almost coinciding with the continuum peak (i.e. near nucleus position); 
3) the (O--Cont) offsets in Dec observed for the molecules of the second group can be explained  by enhanced production in a high-latitude jet-like structure; 
for CS, a 0.8$\arcsec$  offset is obtained with a model including a wide ($\sim 50 ^{\circ}$) conical molecular jet oriented in the plane of the sky and comprising about 25\% of the total production of CS (Fig.~\ref{allmodel}); for SO,  the 1.4$\arcsec$  offset is obtained with a wider jet ($\sim 80^{\circ}$) containing about 40\% of the total production. 
Such  a static wide jet could correspond to the temporal average of a narrower rotating jet; 
the simulations show that a model combining a diurnal asymmetry and a high-latitude jet reproduces qualitatively the line shapes of CS and SO ON--OFF spectra and spectral maps as well as  the (O-Cont) offsets (Fig.~\ref{allmodel}).
4) the position of the centre of brightness in time-integrated maps is not significantly affected by low-latitude rotating gaseous jets, such as the one observed for CO (Sect.~\ref{onoff}).

Fitting in detail the \h2s, CS and SO interferometric maps is challenging, given the complexity of Hale-Bopp coma and lack of temporal coverage.
 This study is beyond the scope of the present paper.
 In Sect.~\ref{results}, we will focuss on the radial distribution of these molecules.  
 
\section{Astrometry: orbital implications}
\label{sect:astro}

It is at this point useful to discuss the Plateau de Bure observations in the context of the orbit determination of the comet. 
 Whether it is or is not necessary to allow for effects of other than gravitational forces on the
Hale-Bopp orbit has long been a rather contentious issue \citep{mar99}. 
 On the one hand, it was difficult to believe that, active though the comet was, the relative mass loss due to outgassing from a nucleus as large as Hale-Bopp was estimated to be  (from 45 to perhaps 70 km diameter, \citealt{altenhoff}, \citealt{weav99}) would be sufficient to produce a detectable effect on the orbit.  
On the other hand, as the timespan of the astrometric observations increased well beyond the comet's perihelion passage, the manner in which the orbital solutions satisfied the data was becoming less and less satisfactory, even when special weighting  schemes were introduced to stress some observations over others.
Eventually, when the observations extended some ten months after perihelion passage, there was no other possibility but to incorporate nongravitational parameters into the solution, including in particular a radial force component outward from the Sun that had a formal mean error of only some three percent.
In broad terms, later orbital solutions, whether computed at the Minor Planet Center or at the Jet Propulsion Laboratory (e.g., MPC 54799 and JPL 220 orbit used in Sect~\ref{intobs}, which both include observations made into Aug. 2005), seemingly confirmed this finding. 

 If there were a significant departure between the centre of mass and the centre of brightness when the comet was most active, it makes sense to attempt entirely gravitational orbital solutions that exclude consideration of the optical astrometry during that time period. 
 Although the last observations were made more than 12 years after a fortuitous prediscovery observation, some 19 percent of the total of 3565 observations in the Minor Planet Center's files were made during the 50 days immediately preceding the comet's perihelion passage. 
 The exclusion of these observations permits a moderately satisfactory gravitational fit to the remainder. 
 The (O--C) residuals in Dec for the omitted timespan are systematically negative to an average extent of about 3$\arcsec$. 
 Only 2 percent of the observations were made during the two months after perihelion passage, but this negative trend also shows during this period of included  observations,  after which there was a six-week gap in the observational record because of the comet's small elongation from the Sun.  
There is likewise a negative trend during the last month of included observations prior to perihelion passage.

\begin{table*}
      \caption[]{J2000.0 orbital elements of C/1995 O1 (Hale-Bopp). Epoch 1997 Mar. 13.0 TT. }
         \label{orbit}
	 \centering
         \begin{tabular}{ l c c c c c c c }
          \hline  \hline
            \noalign{\smallskip}
             Orbit & T & q & Peri & Node & Incli. & e & 1/a \\
                   & 1997 Apr. (TT)  & (AU) & ($^\circ$) & ($^\circ$) & ($^\circ$) & & (AU$^{-1}$) \\
            \noalign{\smallskip}
            \hline
            \noalign{\smallskip}
I &  1.13588 & 0.9141659 & 130.58732 & 282.47062 & 89.43001 & 0.9951316 &  +0.0053255 \\
II & 1.13606 &  0.9141660 & 130.58734 &  282.47061 & 89.43002 & 0.9951315 & +0.0053256 \\
III & 1.13623 & 0.9141657 & 130.58735 & 282.47061 &  89.43003 & 0.9951313 & +0.0053258 \\
 &  ($\pm$ 0.00003)$^{\mathrm{a}}$ & ($\pm$ 0.0000002)$^{\mathrm{a}}$ & ($\pm$ 0.00001)$^{\mathrm{a}}$ & ($\pm$ 0.00001)$^{\mathrm{a}}$ & ($\pm$ 0.00001)$^{\mathrm{a}}$ & ($\pm$ 0.0000002)$^{\mathrm{a}}$ & ($\pm$ 0.0000002)$^{\mathrm{a}}$ \\ 
	    \noalign{\smallskip}
            \hline
         \end{tabular}
\begin{list}{}{}
\item[$^{\mathrm{a}}$] Errorbars for orbit III.
\end{list} 
   \end{table*}

Orbit I in Table~\ref{orbit} was therefore computed from 2693 observations covering the time periods 1995 July 24--1997 Jan. 10 and 1997 July 14--2005 Aug. 4.
Observations with a residual in RA or Dec or both of 2.0$\arcsec$ or larger were also excluded from this gravitational solution.  
The mean residual of the included data is 0.86$\arcsec$, and there are no obvious systematic trends.  
The single prediscovery observation on 1993 Apr. 27 has an (O--C) Dec residual of -2.0$\arcsec$, and the negative trend during the omitted half-year is much as before.    

\begin{table*}
      \caption[]{Astrometry of C/1995 O1 (Hale-Bopp) from continuum observations at 
Plateau de Bure interferometer and residuals with respect to orbits in Table ~\ref{orbit}.}
         \label{pos}
	 \centering
         \begin{tabular}{ c c c c c c c}
          \hline  \hline
            \noalign{\smallskip}
\multicolumn{2}{c}{1997 Mar. UT} & RA (J2000.0) & Dec (J2000.0) & (O$-$C) Orbit I & (O$-$C) Orbit II & (O$-$C) Orbit III \\
\multicolumn{4}{c}{} & ($\delta$RA,$\delta$Dec)
& ($\delta$RA,$\delta$Dec) & ($\delta$RA,$\delta$Dec) \\
day &  h:min:s & h:min:s & $^{\circ}$:$\arcmin$:$\arcsec$ & ($\arcsec$,$\arcsec$) & ($\arcsec$,$\arcsec$) & ($\arcsec$,$\arcsec$) \\
\hline
 09 & 05:00:00 &   22:14:44.864 & +39:22:44:07 & ($-$1.4,$-$0.6) &  ($-$0.9,$-$0.4) & ($-$0.5,$-$0.1) \\ 
09 & 08:00:00  &  22:15:40.688  & +39:27:56.90   &  ($-$1.2,$-$0.6)  &  ($-$0.8,$-$0.4) &   ($-$0.3,$-$0.1) \\
09 & 10:00:00  &  22:16:18.031  & +39:31:24.84  & ($-$1.2,$-$0.6)  &  ($-$0.7,$-$0.4)  &  ($-$0.2,$-$0.1) \\
09 & 12:00:00  &  22:16:55.459  & +39:34:52.18  & ($-$1.4,$-$0.7)  &  ($-$1.0,$-$0.5)  &  ($-$0.5,$-$0.3) \\
11 & 06:00:00  &  22:30:26.714  & +40:45:22.83  & ($-$1.3,$-$0.6)  &  ($-$0.8,$-$0.5)  &  ($-$0.3,$-$0.2)\\
11 & 06:00:00  &  22:30:26.724  & +40:45:23.05  & ($-$1.2,$-$0.4)  &  ($-$0.7,$-$0.2) &   ($-$0.2,\phantom{$-$}0.0)\\
13 & 08:00:00  &  22:47:34.109  & +42:03:01.56  & ($-$1.3,$-$0.5)   & ($-$0.7,$-$0.4) &   ($-$0.2,$-$0.2)\\
13 & 15:00:00  &  22:50:03.197  & +42:13:15.05  & ($-$1.3,$-$0.5)   & ($-$0.7,$-$0.3) &   ($-$0.2,$-$0.2)\\
16 & 16:00:00  &  23:17:10.406  & +43:48:28.38  & ($-$1.1,$-$0.4)   & ($-$0.5,$-$0.4) &    (\phantom{$-$}0.0,$-$0.2)\\
16 & 16:00:00  &  23:17:10.407  & +43:48:28.42  & ($-$1.1,$-$0.4)   & ($-$0.5,$-$0.3) &   (+0.1,$-$0.2)\\
 \noalign{\smallskip}
            \hline
         \end{tabular}
\begin{list}{}{}
\item[$^{\mathrm{a}}$] Continuum geocentric positions from \cite{altenhoff} converted from apparent to J2000.0 coordinates.
\end{list} 
  \end{table*}

     The ten IRAM geocentric apparent positions of the peak of the continuum from \cite{altenhoff} are given in Table~\ref{pos}. 
The first set of residuals shown are from Orbit I, the 2693-observation orbit mentionned above. 
Although these residuals are systematically negative, the important point is that those in declination are only slightly negative, suggesting that the peak of IRAM continuum emission is substantially closer to the comet's centre of mass than the optical astrometric observations are  during this time. 
 The second set of residuals are from an orbit solution (Orbit II in Table~\ref{orbit}) that also includes these IRAM observations.
The consistency of the IRAM data suggests it may be appropriate to weight them more heavily in the solution, specifically, by a factor of five. 
 At the same time, it is reasonable to accept in the solution only residuals in either coordinate that are no more than 1.5$\arcsec$. 
 The resulting orbital elements (Orbit III)  are given in Table~\ref{orbit}.
These satisfy 2249 observations with mean residual 0.71$\arcsec$.  
The residuals of the 1993 and 2005 observations are changed from those of the previous solution by no more than 0.1$\arcsec$. 
 The IRAM residuals from Orbit III are now very commendable.  

   We conclude that there is no need to invoke the existence of nongravitational forces acting on this comet.  
The ``original'' and ``future'' barycentric values of 1/a are +0.003801 and +0.005571 AU$^{-1}$, respectively, corresponding to periods of 4267 and 2405 years.
    
\section{Modelling observations}
\label{model}

In order to press on with the data analysis, we have developed a model able to
simulate both interferometric and \oo{} observations of a synthetic coma. In a first
step, the brightness distribution of molecular lines in the plane of the sky is 
computed using a radiative transfer model. 
This brightness distribution is used in the second step to compute synthetic \oo~line
profiles and to simulate velocity-resolved interferometric observations.   

\subsection{Radiative transfer model}

Computing the brightness distribution of a molecular rotational line in the plane of the sky is a classical radiative transfer problem that can be easily solved out by numerical integration provided we know:
 1) the population distribution of the involved  rotational levels; 
2) the spatial and velocity distribution of the studied species;
 3) the velocity dispersion, assumed here to be thermal at local temperature; 
and  4) the background emission, taken to be the 2.7 K cosmic background. 
It also requires an appropriate choice of number of cells and of cell size, both spatially and spectrally,  in order to get sufficient accuracy.    
Detailed formulae of the cells contribution in the case of an optically thin coma can be  found in Bockel\'ee-Morvan et al. ({\it in preparation}).
 The code used in this paper is more general as it takes into account opacity effects: in addition to the proper emission of each cell (at given Doppler velocity interval) it computes the cell attenuation of the  signal emitted from background cells along the line of sight. 
The output of the programme is a  spectral cube which provides the brightness distribution $F_i(x,y)$ in the plane of the sky ($x$,$y$) for a number of velocity channels $i$.
The computed map is 256$\times$256 pixels wide, with pixels of $0.5 \arcsec \times 0.5 \arcsec$, and velocity  channels of 0.10 km s$^{-1}$. 

The level populations in the cells were computed according to the excitation model of \citet{biver99}. 
The model takes into account collisions with water molecules and electrons which control excitation in the inner coma: we used a total cross-section  for de-excitation by neutral collisions  of  $1-2 \times 10^{-14}$ cm$^2$, and a $x_{ne}$ scaling  parameter equal to 1 for the electron density. 
Radiative excitation by the  Sun radiation is taken into account for CS (\citealt{biver99}).
Radiative excitation is not considered for H$_2$S and SO molecules.
 As detailed in \citet{cro91} and \citet{biver06}, given their short photodissociation lifetime, they do not undergo significant radiative excitation before being dissociated.
 This is  particularly relevant in March 1997   for comet Hale-Bopp, the collisional region of which was large at this time.  
The excitation model takes also into account radiation trapping, which main effect is to increase the size of the region where molecules are at thermal equilibrium. 

\begin{figure}
\resizebox{\hsize}{!}{\includegraphics[angle=-90]{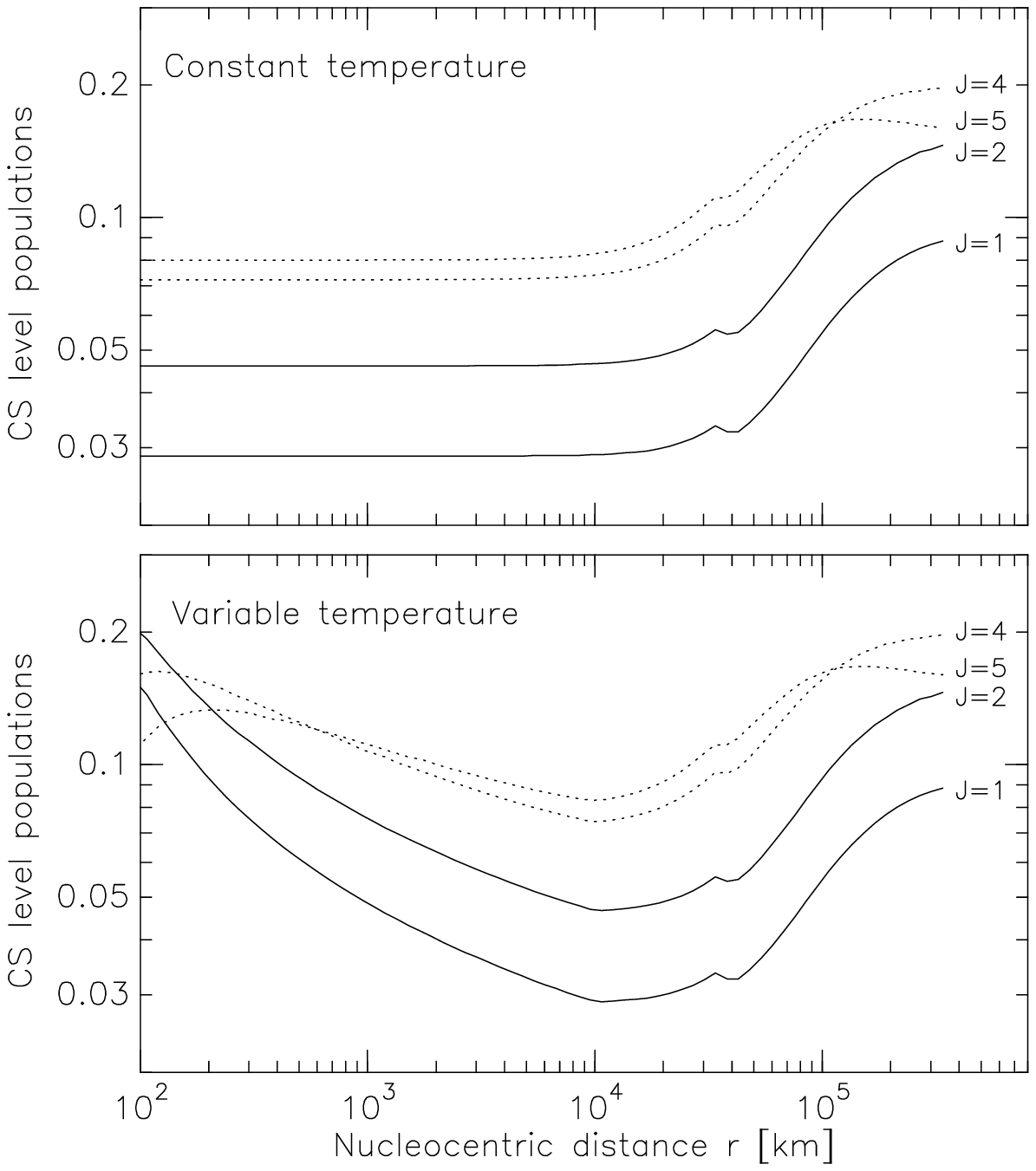}}
\caption{ Population of the rotational levels of CS related to observed transitions: J = 1, 2, 4 and 5 as computed with the excitation model.
 The upper panel corresponds to a constant temperature, the lower to a variable temperature.}
\label{popcs}
\end{figure}

\begin{figure}
\resizebox{\hsize}{!}{\includegraphics[angle=-90]{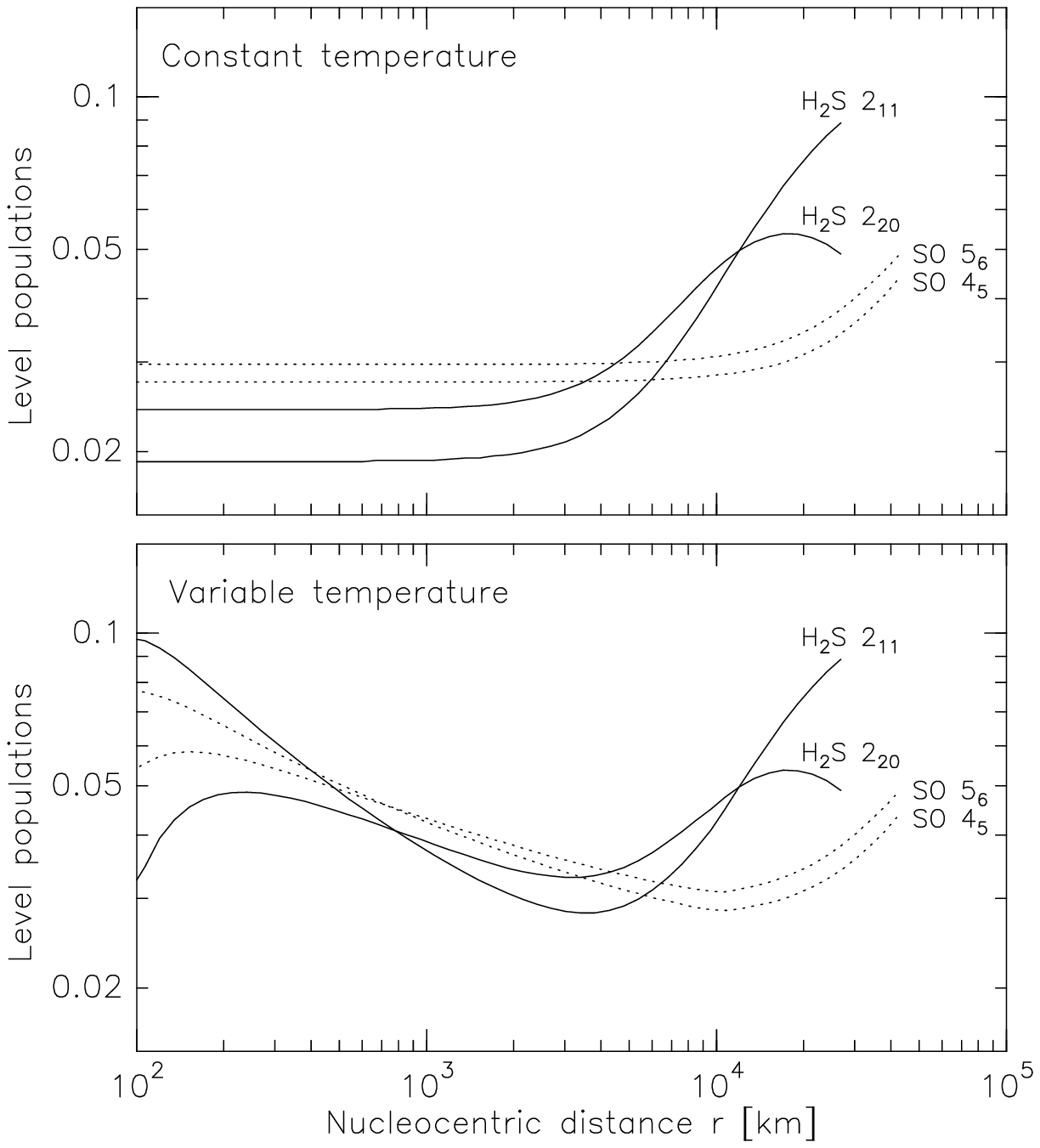}}
\caption{ Population of the \h2s{} and SO rotational levels related to observed transitions  as computed with the excitation model. The upper panel corresponds to a constant temperature, the lower to a variable temperature.}
\label{popsoh2s}
\end{figure}
         
Because the field of view of the Plateau de Bure observations was essentially sampling molecules with rotational level populations at thermal equilibrium, the temperature of the gas is an important parameter of the model.
Both radio (single-dish) (\citealt{biver02}) and infrared (e.g., \citealt{mag99}; \citealt{disanti01}) observations of comet Hale-Bopp in March 1997  suggested gas temperatures $\sim$ 120 K at radial distances of $\sim$10000 km from the nucleus.
 Evidence for a temperature increase with distance to nucleus is reported from long-slit spectroscopy in the infrared (\citealt{mag99}; \citealt{disanti01}). 
Such an increase is predicted in coma hydrodynamic simulations as a result of  photolytic heating (\citealt{combi99}). 
Therefore, we investigated two temperature laws in our model: 
\begin{itemize}
\item \emph{a constant temperature:} $T = 120$ K, derived from the heliocentric dependence of $T$ in comet Hale-Bopp published by \citet{biver02}.
\item \emph{a variable temperature:} $T = 20$ K for $r < 100$ km, $T = 20+50 \log \, (\frac{r \mathrm{[km]}}{100})$ for 100 km  $< r <$ 10000 km and $T = 120$ K for $ r > 10000$ km, where $r$ is the radial distance from the nucleus.
\end{itemize}

The variable temperature law provides a temperature of 70 K at 1000 km from the  nucleus, consistent with the model of \citet{combi99}, but somewhat below the rotational temperatures of 80--90 K derived from infrared measurements with 1$\arcsec$ (1000 km) aperture (\citealt{mag99};  \citealt{disanti01}). 
Therefore, this temperature law allows us to investigate extreme temperature gradients.
The population of the upper and lower levels of the CS, \h2s and SO observed transitions are shown in Figs.~\ref{popcs}--\ref{popsoh2s} for the two temperature laws. 
As the distance from the nucleus increases, the rotational population distributions of the molecules evolve from thermal equilibrium to a cold fluorescence equilibrium, because of declining collision rates.
 This explains the increase in the outer coma of the population   of the levels involved in the observed rotational transitions.
With this temperature law, the modelled \oo~observations (10000--25000 km field of view radius) mostly probe collisionally excited molecules at $T$ = 120 K, while the  interferometric beam (1000 km equivalent radius) is sensitive to colder regions at $T \sim$ 70 K. 
Colder gas in the inner coma results in larger opacity effects, as it will be shown in Sect.~\ref{results}.

The radiative transfer code can accommodate any 3-D density and velocity  distribution as inputs.
 For the purpose of Sect.~\ref{results}, namely the analysis of the radial distribution of CS, \h2s{} and SO, the density distribution was described by the Haser formulas for parent and/or daughter species. 
Since the \h2s{} coma does not show up strong spatial and spectral asymmetries,  the assumption of isotropic outgassing is satisfactory for analysing the \h2s{} data. 
This is less true for SO and CS (Sect~\ref{results}), so that this  must be kept in mind in the discussion of the results.    
We can note that the Haser formula for daughter molecules is appropriate for describing the density distribution of SO and CS in comet Hale-Bopp within the IRAM instrumental field of view given the large size ($>$ 10$^5$ km, \citealt{combi99}) of the collision sphere  in March 1997 compared to the scalelengths of their presumed parents (see
Sect.~\ref{results}). 
Numerical simulations of asymmetric comas discussed in Sect.\ref{intobs} were performed using a combination of conical structures, each of them characterized by a production rate by unit of solid angle.
    
For the gas expansion velocity, we assumed a value about  1.03 $\mathrm{km~s}^{-1} $, derived from the heliocentric dependence of the gas expansion velocity measured in comet Hale-Bopp by \citet{biver02}. 
Other relevant parameters in the Haser model that will be constrained by the observations are the photodissociation rates of CS, \h2s, SO and SO parent, and the CS, \h2s and SO production rates.  

\begin{table}
      \caption[]{Published photodissociation rates at 1 AU from the Sun.}
         \label{beta}
	 \centering
         \begin{tabular}{ l c l }
          \hline  \hline
            \noalign{\smallskip}
             Molecule & $\beta$ [s$^{-1}$] & Reference \\
            \noalign{\smallskip}
            \hline
            \noalign{\smallskip}
\h2s             & $ 2.5 \times 10^{-4} \phantom{\, ^{\star}} $ & Crovisier et al. 1991 \\
CS\phantom{$_2$} & $ 1.0 \times 10^{-5} \phantom{\, ^{\star}} $ & Jackson et al. 1982 \\
                 & $ 2.0 \times 10^{-5} \phantom{\, ^{\star}} $ & Biver et al. 2003 \\
	         & $ 1.0 \times 10^{-4} \phantom{\, ^{\star}} $ & Snyder et al. 2001 \\            
CS$_2$           & $ 1.0 \times 10^{-3} \phantom{\, ^{\star}} $ & Feldman et al. 1999 \\
                 & $ 1.7 \times 10^{-3} \phantom{\, ^{\star}} $ & Jackson et al. 1986 \\ 
SO\phantom{$_2$} & $ 1.5 \times 10^{-4} \phantom{\, ^{\star}} $ & Kim \& A'Hearn 1991 \\
                 & $ 4.9 \times 10^{-4} \phantom{\, ^{\star}} $ & Summers \& Strobel 1996\\ 
                 & $ 6.2 \times 10^{-4} \phantom{\, ^{\star}} $ & Huebner et al. 1992 \\
SO$_{2}$         & $ 2.1 \times 10^{-4} \phantom{\, ^{\star}} $ &  Kim \& A'Hearn 1991 \\
	         & $ 2.1 \times 10^{-4} \phantom{\, ^{\star}} $ & Huebner et al. 1992 \\
	         & $ 2.3 \times 10^{-4} \phantom{\, ^{\star}} $ & Kumar 1982 \\     
	         & $ 2.9 \times 10^{-4} \, ^{\star} $ & Summers \& Strobel 1996\\ 
	    \noalign{\smallskip}
            \hline
 \noalign{\smallskip}
         \end{tabular}

$^{\star}$ The solar spectrum model used pertains to solar maximum.
   \end{table}

\subsection{\oo{} synthetic spectra}

Synthetic \oo{} line profiles are readily obtained from the spectral cube by convolving the brightness distribution obtained for each velocity channel with the primary beam pattern. 
The beam pattern $A(x,y)$ was described by  a 2D-gaussian which width at half maximum is equal to the HPBW given in Table~\ref{log}. 

\subsection{Synthetic interferometric observations}

The simulation of interferometric observations is more complex, as it requires the Fourier transform of the brighness distribution taking into account the \emph{uv--coverage} (see Sect.~\ref{interf}). 

For each channel $i$, the visibilities are defined by \citep[see e.g.,][]{thompson}~:
\begin{equation}\label{def-visibility}
\mathcal{V}_i(\vec{\sigma}) = \frac{c}{\nu \delta v} \int_{4\pi} A(\vec{s}) F_i(\vec{s}) 
\exp(-\frac{2i\pi\nu}{c} \vec{\sigma} \cdot \vec{s}) d\Omega,
\end{equation}
where $\vec{\sigma}$ is the baseline vector for two antennas, with coordinates ($u$,$v$) in the $uv$-plane. 
\vec{s} is a vector in the sky plane which coordinates are $(x,y)$ in radian units.
 $A$ is the power pattern of the  antennas, and $d\Omega$ is an element of solid angle on the sky. 
Unit of the  brightness distribution $F_i(x,y)$ provided by the radiative transfer model is [W m$^{-2}$ sr$^{-1}$]. 
$\mathcal{V}_i(\vec{\sigma})$ is here in units of [W m$^{-2}$ Hz$^{-1}$] or janskys.
Equation~\ref{def-visibility} can be approximated to~:
\begin{eqnarray}
\label{vis}
\mathcal{V}_i(u,v) & = & \frac{c}{\nu \delta v} \int_{-\infty}^{+\infty} \int_{-\infty}^{+\infty} 
A(x,y) F_i(x,y) \nonumber \\
 & & \times \ \exp(-\frac{2i\pi\nu}{c}(ux+vy))\, dx dy.
\end{eqnarray}

The calculation of the visibilities is carried out using the  GILDAS software collection  provided by
IRAM. 
The first step consists in converting the brightness spectral map $F_i(x,y)$ times the primary beam gaussian pattern $A(x,y)$ into GILDAS
format. 
Then a GILDAS task computes the Fourier transform according to  Eq.~\ref{vis}. 
In order to simulate data comparable to observations, it is necessary to take into account the \emph{uv--plane} sampled by the observations. 
In practice, the  GILDAS task reads the $uv$-coverage in the observed data file and computes the $\mathcal{V}_i(u,v)$ for each ($u$,$v$) point. 
This step provides a $uv$-table that can be handled exactly like that recorded during the observations. 
Modelled and observed maps obtained by inverse Fourier Transform can be compared. 

It is also possible to compare the visibilities in the Fourier space. 
Because of symmetry properties  of the Fourier Transform, a radially symmetric brightness distribution results in visibilities $\mathcal{V}_i(u,v)$ that only depend upon the $uv$-radius $\sigma$ =$\sqrt{u^2+v^2}$. 
Observed visibilities that have been radially averaged in the $uv$-plane can be  directly compared to modelled  $\mathcal{V}_i(\sigma)$. 
In such a way, the  mean (angularly averaged) radial brightness distribution of observed species can be investigated. 
For this study, visibilities integrated over velocity  $\sum_{i}\mathcal{V}_{i}(\sigma) \delta v$ [Jy km s$^{-1}$] (with $\delta v$ equal to channel width) will be compared (Sect.~\ref{results}). 
Note that the line flux measured in \oo~mode corresponds to the visibility at  $\sigma$ = 0.

\section{Analysis of the radial distribution of CS, \h2s{} and SO}
\label{results}

Two different methods are used to analyse the  radial distribution of CS, \h2s{} and SO.
Let us define $F_{\rm SD}$ as the line area of the \oo~spectra and  $F_{\rm Int}$ as the line area at maximum intensity in the interferometric maps, which observed values are given in Table~\ref{log}.
 Because of the factor of $\sim$ 10 difference in angular resolution between \oo~and interferometric observations, the flux ratio $R = F_{\rm SD}/F_{\rm Int}$ is strongly sensitive to the spatial distribution. 
 We can thus compare observed and modelled flux ratios to constrain model parameters. 
The more extended the brightness distribution is, the larger the flux ratio $R$.
In the second method, we compare observed and modelled visibilities as a function of $uv$-radius, as explained previously, including the \oo~data point. 

Input free parameters in the model are photodissociation and  production rates. 
The production rate  $Q$ is adjusted iteratively by running the model several times in order to match at best the flux ratio and the observed intensities $F_{\rm SD}$ and $F_{\rm Int}$. 
Indeed, the intensities (particularly $F_{\rm Int}$) are affected by small but significant  opacity effects, so that the flux ratio depends on $Q$.             

\subsection{Hydrogen sulfide \h2s}
\begin{table}
    \caption[]{Observed and modelled fluxes  (in Jy km s$^{-1}$) and flux 
ratios.}
         \label{flh2s}
         \centering
	 \begin{tabular}{ c c r r r r }
            \hline\hline
            \noalign{\smallskip}
	     & Observations$^{a}$ &  \multicolumn{4}{c}{  Models$^{b}$} \\
             \cline{3-6}\\
            &  & \multicolumn{2}{c}{ $T =120 \,  \mathrm{K}$} & \multicolumn{2}{c}{  $T$ variable} \\
            & & $Thin$ & $Thick$ & $Thin$ & $Thick$ \\
            \noalign{\smallskip}
	    \hline
            \noalign{\smallskip}
            \multicolumn{2}{l}{H$_2$S $2_{20}$--$2_{11}$}& & & &\\
	    $F_{SD}$                  &  $18.9 \pm 2.8\phantom{0}$   &19.14	    &17.20    & 19.40      & 16.60    \\
  \noalign{\smallskip}
	    $F_{Int}$                 & $3.51 \pm 0.55$   & 4.89   &	    3.96   & 5.15     & 3.60    \\
  \noalign{\smallskip}
	    $R$ & $5.38 \pm 1.16$   & 3.91    &	    4.34 & 3.76 &   4.61\\
\hline
            \noalign{\smallskip}     
	    \multicolumn{2}{l}{CS $J$(2--1)} & & & &\\ 
	    $F_{SD}$  & $15.2 \pm 2.3\phantom{0}$  &15.1 & 14.3    & 14.6      & 13.7    \\
  \noalign{\smallskip}
	    $F_{Int}$ & $1.47 \pm 0.24$   & 1.07   & 1.02   & 1.50     & 1.29    \\
  \noalign{\smallskip}
	    $R$ & $10.3 \pm 2.3\phantom{0}$  & 14.1 & 14.0 & 9.7 &  10.6\\
\hline
            \noalign{\smallskip}     
	    \multicolumn{2}{l}{CS $J$(5--4)} & & & & \\ 
	    $F_{SD}$ &  $\phantom{.}213 \pm 32\phantom{.0}$  & 231 & 213    & 227     & 201    \\
  \noalign{\smallskip}
	    $F_{Int}$ & $5.4 \pm 1.1$  & 11.4   & 8.22   & 14.7     & 7.30    \\
  \noalign{\smallskip}
	    $R$ & $\phantom{.0}39 \pm 10\phantom{.0}$ & 20.3 & 25.9 & 15.4 &  27.5\\

	    \noalign{\smallskip}

	    \noalign{\smallskip}
            \hline
         \end{tabular}
\begin{list}{}{}
\item[$^{a}$] Line integrated intensity in \oo{} mode   ($F_{\rm SD}$) and at the centre of the interferometric map ($F_{\rm Int}$)  and their ratio $R = F_{\rm SD}/F_{\rm Int}$. 
Unlike in Table~\ref{log},   error bars take into account a 15\% uncertainty in the calibration. 
\item[$^{b}$] Model calculations of $F_{\rm SD}$ and $F_{\rm Int}$, assuming a  constant or variable temperature in the coma (see Sect.~\ref{model}). 
  Results labelled as $thick$ (respectively $thin$) indicate whether opacity effects were included or not in the calculations. 
Calculations were performed with $\beta_{\rm H_{2}S} = 2.5 \, \times \, 10^{-4} \, s^{-1}$, $\beta_{\rm CS_2}$ = 1.7 $\times$ 10$^{-3}$ s$^{-1}$, $\beta_{\rm CS}$ = 2 $\times$ 10$^{-5}$ s$^{-1}$.
 The assumed production rates are $Q_{\rm H_2S}$ = 1.35 $\times$ 10$^{29}$ s$^{-1}$, $Q_{\rm CS}$ = 2.0 $\times$ 10$^{28}$ s$^{-1}$ (respectively $Q_{\rm H_2S}$ = 1.2 $\times$ 10$^{29}$ s$^{-1}$, $Q_{\rm CS}$ = 1.8 $\times$ 10$^{28}$ s$^{-1}$) for $T$ = 120 K (respectively, variable $T$) law. 
\end{list}
\end{table}

The photodissociation rate of \h2s{} has been determined by \citet{cro91} using laboratory measurements (Table~\ref{beta}). 
The resulting \h2s{} scalelength for comet Hale-Bopp in March 1997 is $\sim$ 3900 km, i.e. intermediate between the angular resolution of \oo~and interferometric observations. 
Therefore, the flux  ratio $R$ is expected to reflect, in large part, \h2s{} photolysis.    

The observed flux ratio $R$ can be reproduced within \hbox{1-$\sigma$} with the model for a photodissociation rate of \h2s{} in the range $1.5-3 \times 10^{-4} \, \mathrm{s}^{-1}$. 
This range includes the value from \citet{cro91}, for which computed flux densities and flux ratios $R$  are given in Table \ref{flh2s}, for the two temperature laws, together with measured values. 
Calculations in the assumption of an optically thin line are also given.  
The good agreement between model  and observations (within 1-$\sigma$) in the optically $thick$ case  shows that the  radial extension of \h2s in the coma is well explained by direct release of \h2s from the  nucleus.  
Opacity effects significantly affect the line flux in the interferometric beam: by 23\% with the constant temperature law to 40\% for $T$ variable. 
The retrieved H$_2$S production rates are 1.35 and $1.2 \times 10^{29}$ s$^{-1}$, for the constant and variable $T$ laws, respectively. 

\begin{figure}
\resizebox{\hsize}{!}{\includegraphics[angle=-90]{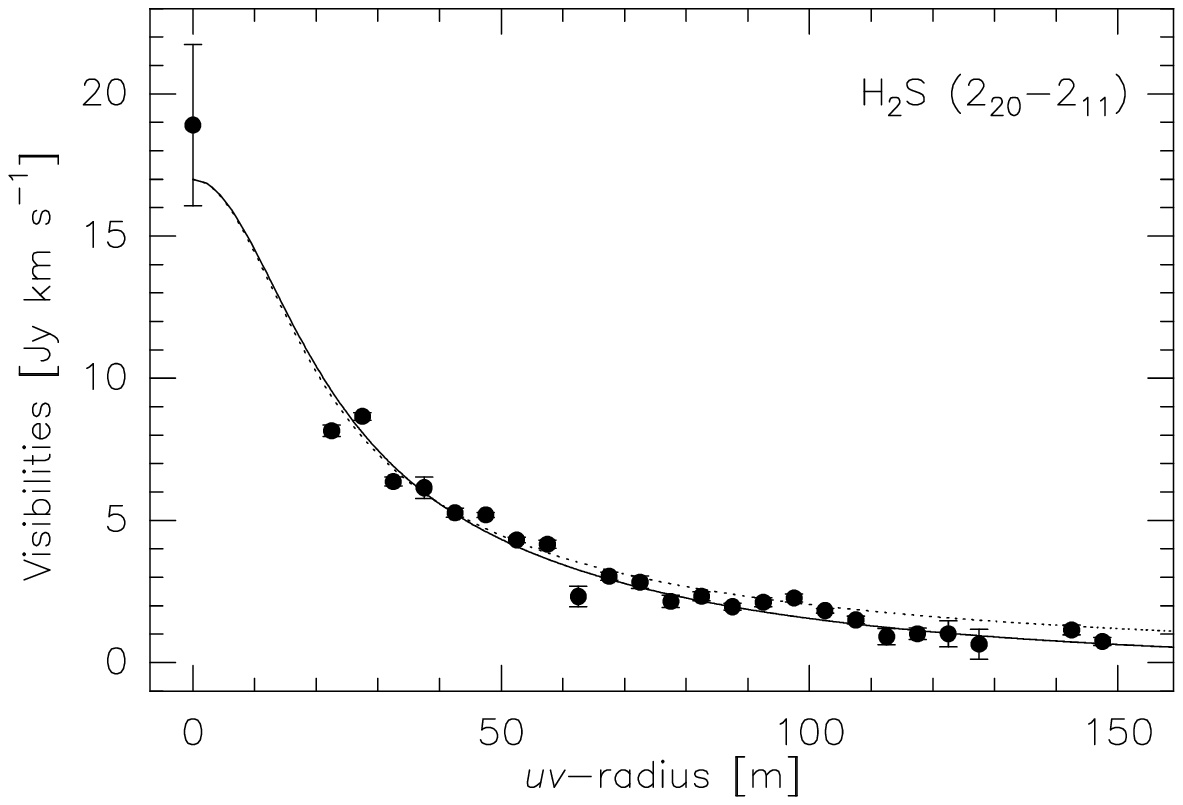}}
\caption{Amplitude of the visibility of \h2s{} 217 GHz line versus $uv$-radius. 
Circles with error bars: observations. 
Dotted lines: model assuming constant $T$. 
Thick lines: model assuming variable $T$. 
In both cases, calculations were performed with $\beta_{\rm H_{2}S} = 2.5  \times  10^{-4} \, \mathrm{s}^{-1}$.}
\label{visih2s}
\end{figure}

Figure~\ref{visih2s} shows the comparison between modelled (with opacity effects included) and measured visibilities for the two temperature laws. 
Both models reproduce correctly the  radial evolution of the visibilities. 
At large $uv$-radii, the fit is better with the  model assuming a variable temperature.
Fluctuations apart the model fit trace spatial asymmetries in the coma. 

The \oo~\h2s spectrum given by the model is superimposed on the observed spectrum in Fig~\ref{spetot}. 
The agreement is satisfactory, not considering the small spectral asymmetry in the observed spectrum which is due to non-isotropic \h2s density and velocity distribution as discussed in Sect.~\ref{obs}.   

\subsection{Carbon sulfide CS}  
\begin{figure}
\resizebox{\hsize}{!}{\includegraphics[angle=-90]{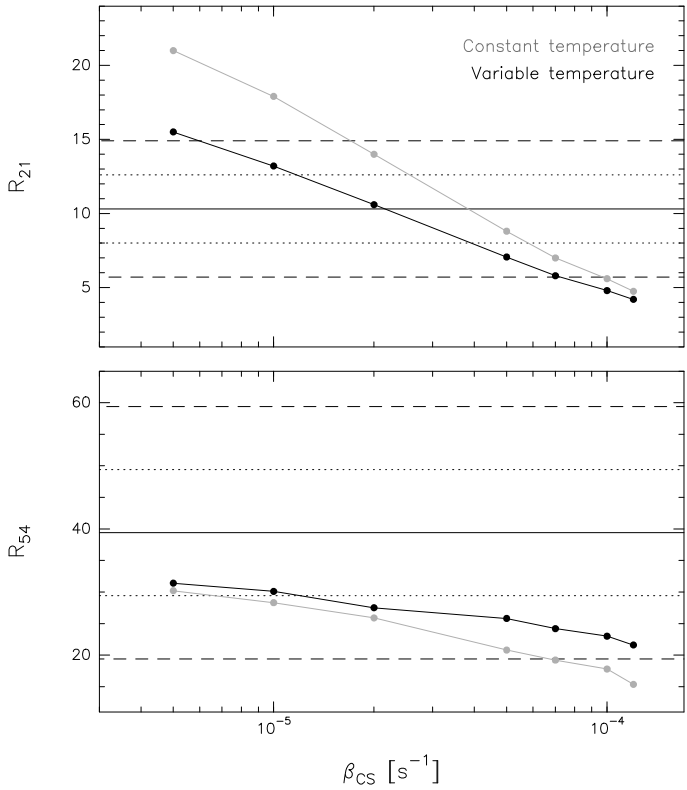}}
\caption{Modelled flux ratios $R$ as a function of CS photodissociation rate for CS $J$(2--1) ($R_{21}$, top) and $J$(5--4) ($R_{54}$, bottom) lines  for $\beta_{\rm CS_2}$ =  1.7 $\times$ 10$^{-3}$ s$^{-1}$ (value at $r_h$ = 1 AU). 
The dotted (respectively dashed) lines show the 1-$\sigma$ (resp. 2-$\sigma$) range around the observed value (continuous line).}
\label{mes_cs}
\end{figure}

\begin{figure}
\resizebox{\hsize}{!}{\includegraphics[angle=-90]{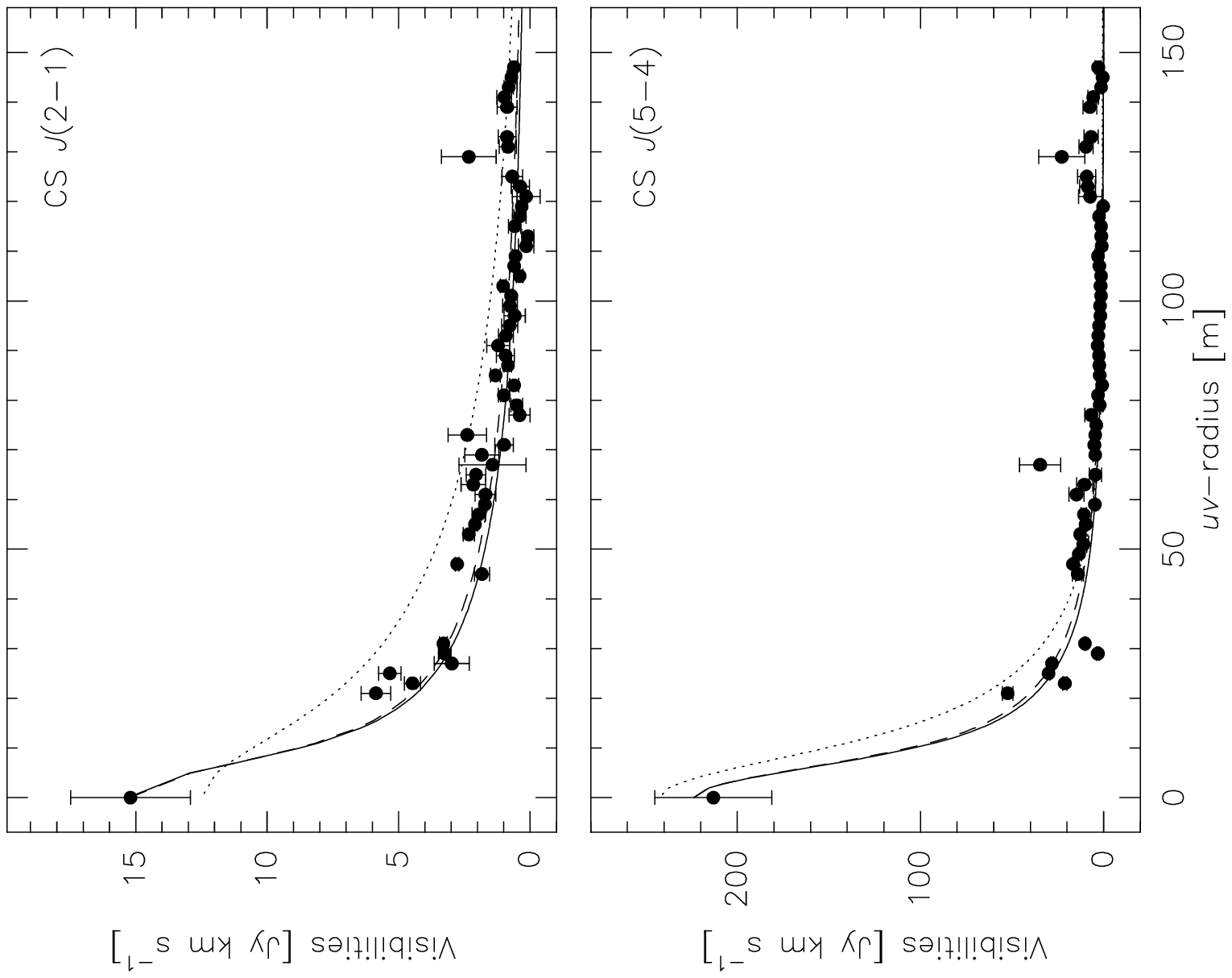}}
\caption{Amplitude of the visibility as a function of $uv$-radius  for CS $J$(2--1) (upper panel) and CS $J$(5--4) (lower panel). 
Measurements are shown by dots with error bars. Displayed modelled  visibilities are for a variable temperature law and parameter sets: 
($\beta_{\rm CS}$~=~2~$\times$~10$^{-5}$~s$^{-1}$, $\beta_{\rm CS_{2}}$~=~1.7~$\times$~10$^{-3}$~s$^{-1}$) in solid line,
 ($\beta_{\rm CS}$~=~2~$\times$~10$^{-5}$~s$^{-1}$, $\beta_{\rm CS_{2}}$~=~1~$\times$~10$^{-3}$~s$^{-1}$) in dashed line and 
($\beta_{\rm CS}$~=~1~$\times$~10$^{-4}$~s$^{-1}$, $\beta_{\rm CS_{2}}$~=~1.7~$\times$~10$^{-3}$~s$^{-1}$) in dotted line.}
\label{visics}
\end{figure}

Ultraviolet observations of CS performed with the {\it International Ultraviolet Explorer} (IUE) have shown that CS$_2$ is the likely parent of CS (\citealt{jack82}, \citeyear{jack86}). 
More recently, a comparison between unidentified lines in comet 122P/de Vico spectra and jet cooled laboratory CS$_2$ spectra suggests that CS$_2$ is indeed present in cometary atmospheres (\citealt{jack04}). 
The CS$_2$ lifetime at 1 AU is estimated between 590 s (\citealt{jack86}) and 1000 s (\citealt{feld99}) from observations and laboratory measurements (i.e., photodissociation rate $\beta_{\rm CS_2}$ between 1 and 1.7 $\times$ 10$^{-3}$ s$^{-1}$).

The photodissociation rate of CS is debated. From IUE observations, \citet{jack82} deduced $\beta_{\rm CS}$ = 1 $\times$ 10$^{-5}$ s$^{-1}$.
 From CS $J$(2--1) interferometric observations of comet Hale-Bopp performed with the BIMA array, \citet{sny01} inferred $\beta_{\rm CS}$ = 1 $\times$ 10$^{-4}$ s$^{-1}$, i.e. a value ten times higher. 
On the other hand, single-dish millimetric observations of comets at small heliocentric distances suggested  $\beta_{\rm CS} \approx 2 \times 10^{-5}$ s$^{-1}$ (\citealt{biver03}).         

The field of view radius  of our CS \oo~observations is 10000 to 25000 km, depending on the line (Table~\ref{log}).
Therefore, the calculated flux ratio $F_{\rm SD}$ and $F_{\rm Int}$ should be strongly dependent upon the assumed CS scalelength, when exploring values between 10000 and 100000 km. 
On the other hand, the  interferometric beams (750 to 2000 km radius, for the $J$(5--4) and $J$(2--1) lines, respectively) are comparable to the CS$_2$ scalelength, which makes the determination of the CS$_2$ lifetime more difficult, given large uncertainties in CS lifetime.

Figure~\ref{mes_cs} shows the dependence of the flux ratio of the $J$(2--1) ($R_{21}$) and  $J$(5--4) ($R_{54}$) lines with the photodissociation rate of CS, assuming a constant or a variable temperature in the coma. 
The CS$_2$  photodissociation rate was taken equal to $\beta_{\rm CS_2}$ =  1.7 $\times$ 10$^{-3}$ s$^{-1}$ (value at $r_h$ = 1 AU). 
The observed ratios can be reproduced within 2-$\sigma$, with  $\beta_{CS}$ in the range  2--7 $\times$ 10$^{-5}$ s$^{-1}$ (case of constant $T$) or with $\beta_{CS}$ between 0.7 and 6 $\times$ 10$^{-5}$ s$^{-1}$ (case of variable $T$). 
The same computations were made with $\beta_{\rm CS_2}$ =  1 $\times$ 10$^{-3}$ s$^{-1}$ (\citealt{feld99}) in order to investigate the sensitivity of the results to this parameter. 
A satisfactory fit within 1-$\sigma$  is obtained for both lines ($R_{21}$ = 12 and  $R_{54}$ = 29) with  $\beta_{\rm CS}$ in the range 2--5 $\times$ 10$^{-5}$ s$^{-1}$ and a variable $T$. 
In summary, our data are consistent with a CS photodissociation rate of 1--5 $\times$ 10$^{-5}$ s$^{-1}$. On the basis of the $J$(2--1) data alone, the value $\beta_{\rm CS}$ = 2--3 $\times$ 10$^{-5}$ s$^{-1}$ is preferred.     
Models assuming a variable temperature in the coma provide the best results.

Table~\ref{flh2s} shows that opacity effects strongly affect the intensity of the $J$(5--4) line in the interferometric   beam, justifying our radiative transfer model. 
They are less significant for  $J$(2--1), and for $T$ = 120 K than for variable $T$. 
These trends can be easily explained: 1) the  $J$(5--4) is intrinsically stronger and observed with a smaller field of view;
 2) with the adopted variable temperature law, the lower rotational levels of the $J$(2--1) and $J$(5--4) transitions  (namely $J$ = 1 and $J$ = 4) are more significantly populated in the inner 
coma (see Fig. \ref{popcs}), hence more significant self-absorption effects. 
The retrieved CS production rate is $\sim$ 2 $\times$ 10$^{28}$ s$^{-1}$ with  $\beta_{\rm CS}$ = 1--2 $\times$ 10$^{-5}$ s$^{-1}$.        

In Fig.~\ref{visics} we show the radial evolution of the visibilities for different model parameters, assuming a variable temperature in the coma.
Models with  $\beta_{\rm CS}$ = 2 $\times$ 10$^{-5}$ s$^{-1}$ and $\beta_{\rm CS_2}$ equal to 1 $\times$ 10$^{-3}$ s$^{-1}$ and  1.7 $\times$ 10$^{-3}$ s$^{-1}$ reproduce correctly the observed visibilities, as expected from the study of the fluxes ratio.
Visibilities computed with $\beta_{\rm CS}$ = 1 $\times$ 10$^{-4}$ s$^{-1}$ do not match the data. 

The high value $\beta_{\rm CS} = 1 \times 10^{-4} \, \textrm{s}^{-1}$ inferred  by \citet{sny01} from comet  Hale-Bopp observations of CS $J$(2--1) with the BIMA array is ruled out by our analysis of Plateau de Bure observations.
\citet{sny01} used a different approach. They analysed CS radial profiles  deduced from the interferometric maps, but did not consider the autocorrelation measurements that they acquired nor the radial evolution of the visibilities in the $uv$--plane. 
However, for  $\beta_{\rm CS} = 1 \times 10^{-4} \, \textrm{s}^{-1}$, their autocorrelation data yield systematically higher (by a factor 2.5--3) CS production rates than the cross-correlation data. 
We computed that, if $\beta_{\rm CS}$ is decreased to  $\sim 3 \times 10^{-5} \, \textrm{s}^{-1}$, the production rates derived by the two sets of data can be reconciled.    

\subsection{Sulfur monoxide SO}
\label{betaso}

\begin{figure}
\resizebox{\hsize}{!}{\includegraphics[angle=-90]{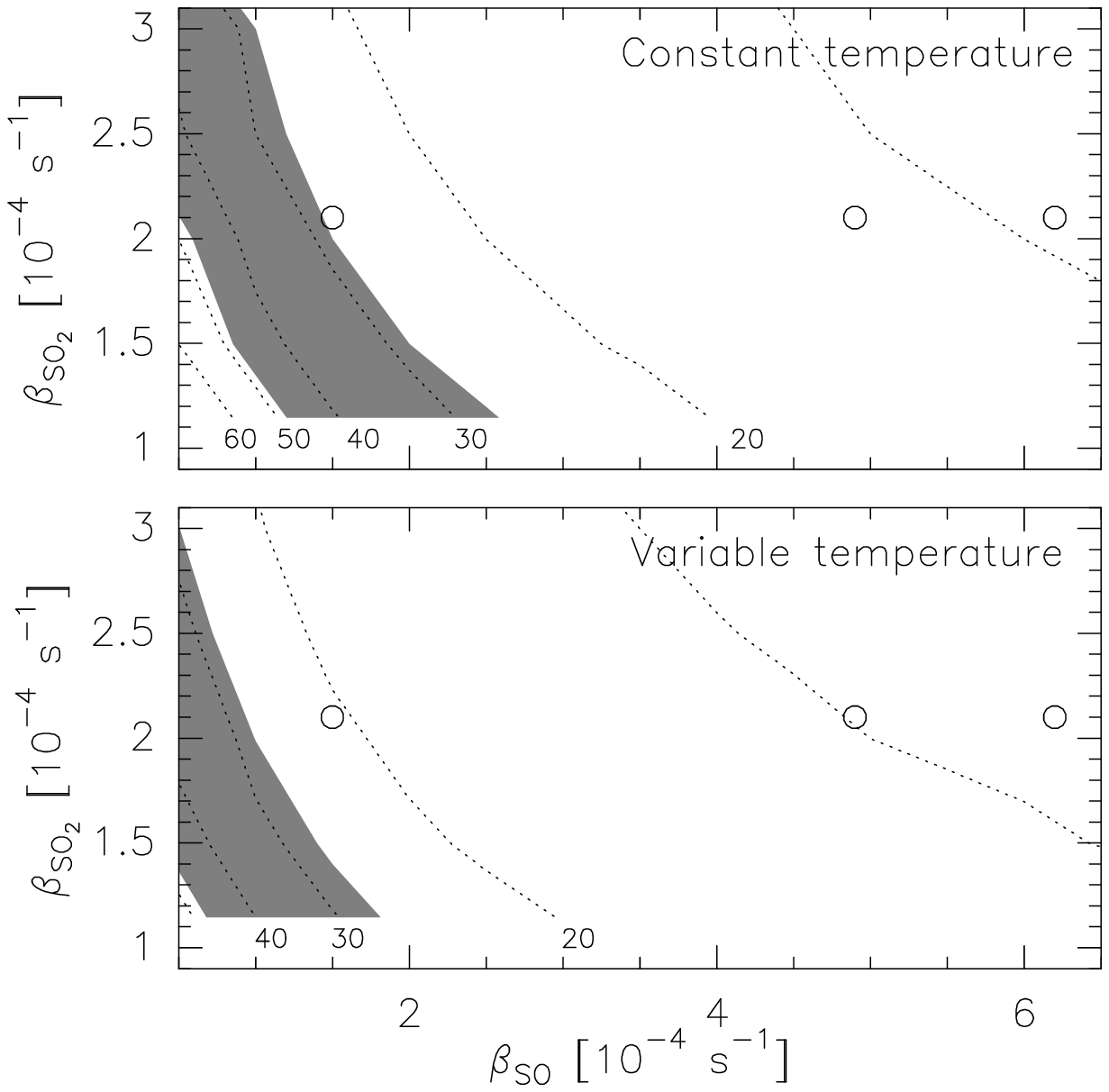}}
\caption{The flux ratio $R = \frac{F_{\rm SD}}{F_{\rm Int}}$ as a function of  $\beta_{SO}$ and $\beta_{SO_{2}}$. 
Upper panel: for constant temperature $T$ = 120 K; 
lower panel: for the variable temperature law.
Contour values are given in the bottom of the figures.
Circles correspond to ($\beta_{SO}$, $\beta_{SO_{2}}$) values published in the literature (Table~\ref{beta}). 
The region where the computed $R$ is in agreement within 1-$\sigma$ with the observations ($R = 38 \pm 10$) is filled in grey.
}
\label{plans}
\end{figure}

As pointed out in Sect.~\ref{intobs},  there is clear evidence that the SO radial brightness distribution is much more extended than that of \h2s.
The flux ratio for the SO line is $R = 38 \pm 10$, taking into account  calibration uncertainties, to be compared to $R = 5.4 \pm 1.2$ for the \h2s line (Table~\ref{flh2s}). SO is a radical and a photodissociation product of SO$_2$, also observed in comet Hale-Bopp (\citealt{dbm2000}). 
Published SO$_2$ photodissociation rates  are in general consistent with each other: $\beta_{\rm SO_{2}} = 2.1 - 2.3  \times  10^{-4} \, \mathrm{s}^{-1}$ for quiet sun and  $2.9  \times  10^{-4} \, \mathrm{s}^{-1}$ for active sun (see Table \ref{beta}).
A significantly lower value was derived by \citet{kum82}, but only one channel of the SO$_{2}$ photodissociation process was considered.
We will not consider this value in this work.

In contrast, published values of $\beta_{\rm SO}$ differ by up to a factor of 4 (Table~\ref{beta}). 
Because SO and \so2 photodissociative scalelengths  are intermediate between the primary and   interferometric fields of view, they can be somewhat constrained by the observations. 
Model calculations were performed for many couples ($\beta_{\rm SO}$, $\beta_{\rm SO_{2}}$)  in the range of published values. Isocontours of the flux ratio $R$ as a function of $\beta_{\rm SO}$ and $\beta_{\rm SO_{2}}$ are presented in Fig.~\ref{plans}.

\begin{figure}
\resizebox{\hsize}{!}{\includegraphics[angle=-90]{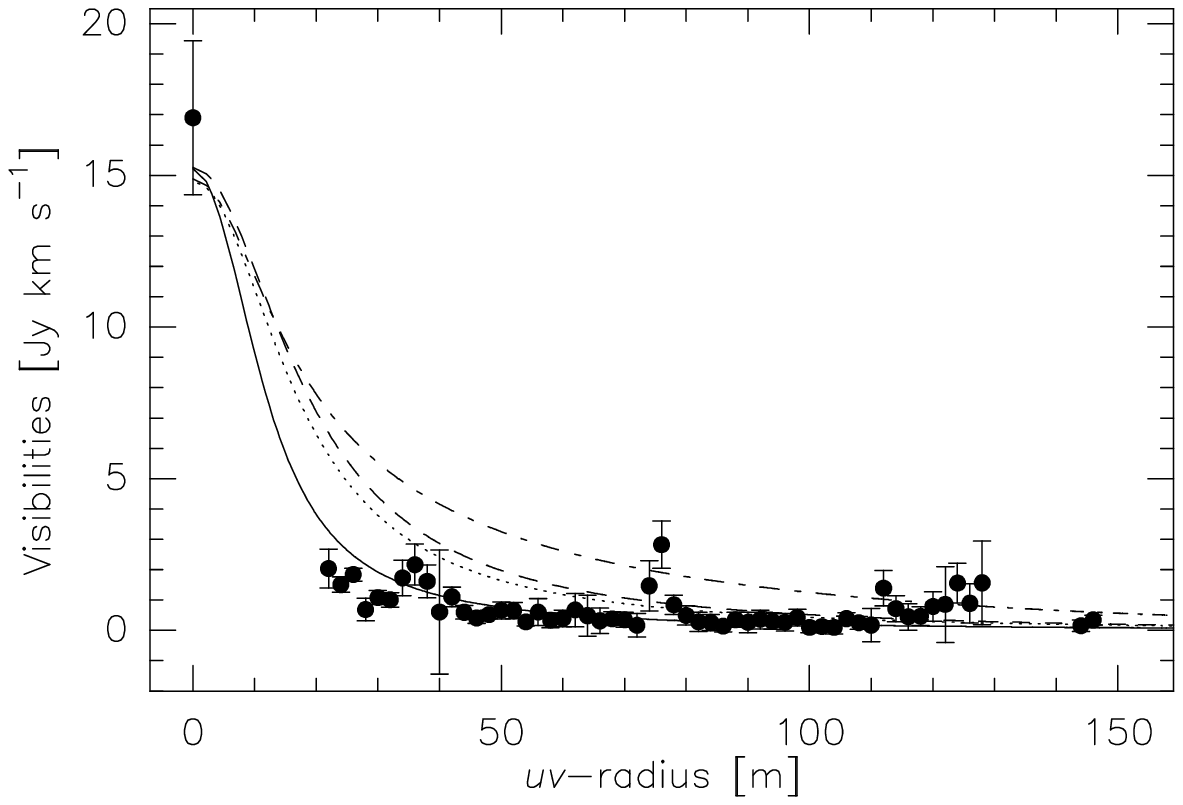}}
\caption{Amplitude of visibility of SO 220 GHz line versus $uv$-radius.
Observations: circles with error bars. 
Continuous, dotted and dashed lines correspond to model calculations assuming that SO is created by the photodissociation of SO$_{2}$ with  $\beta_{SO}$ of 1.5, 4.9 and $6.2 \times 10^{-4} \, \mathrm{s}^{-1}$, respectively. 
The dotted-dashed line tests the case of a pure nuclear origin for SO, with $\beta_{SO} = 1.5 \times 10^{-4} \, \mathrm{s}^{-1}$.  
Calculations correspond to variable $T$ law.
 Similar results are obtained with a constant  $T$ = 120 K.}
\label{visiso}
\end{figure}

In the assumption that SO is produced solely by the photodissociation of \so2, the observations are consistent (within 1--1.5-$\sigma$)  with the published value of $\beta_{\rm SO} = 1.5 \times 10^{-4} \, \mathrm{s}^{-1}$ (\citealt{kim91}), but are not compatible with values as high as $5-6 \times 10^{-4} \, \mathrm{s}^{-1}$ \citep{hue92,summers96}. 
The same conclusion is obtained from the study of the visibility curve (Fig.~\ref{visiso}). 
Direct release of SO from the nucleus is fully excluded (Fig.~\ref{visiso}). 

Published $\beta_{\rm SO}$ rates are all based on the same laboratory data (Phillips, 1981). Discrepancies may be ascribed to the complexity of the laboratory SO absorption spectrum (which shows a series of strong narrow lines over a broad continuum) and the fact that the spectrum is only available in a graphic version.
 From our own analysis of the spectrum, we put a firm lower limit  $\beta_{\rm SO}$ $> 1.5 \times 10^{-4} \, \mathrm{s}^{-1}$  using the lower envelope to the SO spectrum. 
Considering in addition the  narrow lines, we estimate $\beta_{\rm SO}$ = $3.2 \times 10^{-4} \, \mathrm{s}^{-1}$, but this estimation is still uncertain due to the lack of a digital version of the spectrum. 

 How this result could be reconciled with our measurements? 
First, because SO emission is strongly anisotropic, can we exclude a flaw in our analysis due to simple approximations made on the coma structure? 
Calculations performed for an anisotropic coma which reproduces the main characteristics of the SO map (Sect.\ref{intobs}) show that  our analysis should not be significantly affected by the presence of jets in  the coma. 
In addition, a separate reduction of the blue and red velocity parts of the SO spectrum shows that both parts display the same intensity flux ratio $R$ (equal to that measured for the whole emission), indicating that jet structures do not affect the overall evolution of the visibilities with baseline length.

A $\beta_{\rm SO}$ value higher  than that which fits the observations  implies that some process makes the SO brightness radial distribution more extended than expected from SO$_2$ photodissociation. 
We examined several possibilities. Among those that can be excluded are:
1) a $\beta_{\rm SO_2}$ value lower than published values; 
shielding of SO$_2$ photodissociation by water is insufficient (the inner coma where UV shielding by water is significant may be evaluated to $r < 1000$ km); 
2) a kinetic temperature that is much higher ( $\sim$ 200 K) in the inner coma observed by the interferometric beam than in the outer coma  observed by the primary beam, that would be in contradiction with measurements \citep{mag99,disanti01} and hydrodynamical models \citep{combi99}. 
This leaves the possibility that SO$_2$ is not the only source of SO, or that SO$_2$ is itself produced by some extended source in the coma such as grains or complex molecules.          
        
The SO production rate deduced from the observations is 2.4--2.7 $\times$ 10$^{28}$ s$^{-1}$ for $\beta_{\rm SO} = 1.5 \times 10^{-4} \, \mathrm{s}^{-1}$ and  $\beta_{\rm SO_{2}} = 2.1 \times 10^{-4} \, \mathrm{s}^{-1}$. 
Based on the same \oo~data, \citet{dbm2000} inferred a higher value for the SO production rate due to different assumptions on the beam efficiency and kinetic temperature.
SO$_2$ was observed in comet Hale-Bopp on several dates in March and April 1997 using the IRAM Plateau de Bure and 30-m antennas (\citealt{dbm2000}). 
The measurements yield a production ratio  $Q_{\rm SO}$/$Q_{\rm SO_2} \sim 2$ in March 1997. 
The discrepancy increases when a higher $\beta_{\rm SO}$ value is adopted (e.g., $Q_{\rm SO}$/$Q_{\rm SO_2} \sim 6$ for $\beta_{\rm SO} = 6.2 \times 10^{-4} \, \mathrm{s}^{-1}$).   
\citet{dbm2000} suggest that $Q_{\rm SO_2}$ is possibly underestimated in their calculations as they assumed LTE for the SO$_2$ rotational population distribution, and neglect electronic and vibrational excitation. 
Another possible explanation is the presence of another source of SO in the coma. 
As discussed above, this would allow a more satisfactory interpretation of the observed SO radial distribution, given available estimations of $\beta_{\rm SO}$.         

The synthetic SO~\oo~spectrum computed with our isotropic model is plotted in Fig.~\ref{spetot}.
 Though it does not match in details the observed asymmetric SO spectrum, the line width and wings are correctly reproduced.
 This  indicates that, as assumed in the model, the SO radicals which are within the field of view, though daughter products, do indeed partake in  the kinetics of the gas flow. 

\section{Summary}
\label{summary}

Observations carried out in cross- and autocorrelation modes with the IRAM Plateau de Bure interferometer were analysed to study the spatial distribution of sulfur-bearing molecules in the coma of comet Hale-Bopp using a radiative transfer code. 
The radial distribution of  \h2s~is in agreement with a direct release from the nucleus and is well reproduced by our model using the \h2s~photodissociation rate published in the literature, thereby validating our approach.
 
On this basis, we studied the radial distribution of CS emission. 
This radical is believed to be produced by the photolysis of the short-lived CS$_{2}$ molecule. 
The angular resolution of the observations prevented us to obtain observational constraints on the CS$_{2}$ photodissociation rate.
The photodissociation rate of CS is not precisely determined because no satisfactory measurements can be done in the laboratory on this radical.
 Our data are consistent with  $\beta_{CS}$ in the range $1-5 \times 10^{-5} \, \mathrm{s}^{-1}$, which is in agreement with the value of $2 \times 10^{-5} \, \mathrm{s}^{-1}$ proposed by \citet{biver03} but rules out $\beta_{CS} =  1 \times 10^{-4} \,\mathrm{s}^{-1}$ derived by \citet{sny01} from interferometric observations 
with the BIMA array. 
With $\beta_{CS} \sim 3 \times 10^{-5} \, \mathrm{s}^{-1}$, the CS production rates derived from autocorrelation and cross-correlation BIMA data are reconciled.     

SO  was first detected in comet Hale-Bopp, as well as its presumed parent product SO$_{2}$ \citep{dbm2000}. 
Values of $\beta_{SO}$ available in the literature differ by a factor of $\sim$4 between the lowest ($1.5 \times 10^{-4} \, \mathrm{s}^{-1}$, \citealt{kim91}) and the highest value  ($6.2 \times 10^{-4} \, \mathrm{s}^{-1}$, \citealt{hue92}). 
From our own analysis based on the same laboratory absorption spectrum, we estimate $\beta_{SO} \sim 3.2 \times 10^{-4} \, \mathrm{s}^{-1}$.  
In contrast, assuming that SO$_2$ is the parent of SO, the SO interferometric data suggest $\beta_{SO}$ $= 1.5 \times 10^{-4} \, \mathrm{s}^{-1}$. 
The discrepancy between measured and computed values of the SO photodissociation rate may indicate that SO$_{2}$ is not the sole parent of SO, or that SO$_{2}$ is itself created by some extended source in the coma.
Note that the sulfur chemistry in comets is far from being fully understood: the origins of S$_2$ and NS radicals are still unknown \citep{dbm2004}.  
In the future, interferometric mapping of SO and SO$_{2}$ with the Atacama Large Millimeter and submillimeter Array (ALMA) will provide relevant informations about the SO creation process in cometary atmospheres (\citealt{biver05}). 
 Further laboratory experiments on SO absorption in the UV are also strongly encouraged. 

The data show clear evidence for anisotropic spatial distributions in the coma: 
 (1) \oo~spectra are asymmetric;  
(2) the visibilities as a function of $uv$-radius do not follow the expected curves in the case of a 1-D outflow ; 
(3) maps of CS and SO line brightness distributions do not peak at the position of the continuum maps obtained simultaneously which should be close to the nucleus position; 
(4) maps show strong and specific spectral variations in the coma.  
On the basis of model calculations, we show that SO and CS present a wide jet-like structure that may correspond to the gaseous counterpart of the strong high latitude dust jet detected at optical wavelengths.
This structure is not observed for \h2s. 
All three molecules show day/night spatial asymmetries. 
Spectral temporal  variations related to nucleus rotation are only observed for CS. 
The strong CO rotating  jet observed at Plateau de Bure interferometer has no \h2s, CS and SO counterpart. 
These differences in the outgassing patterns of different volatiles in the Hale-Bopp coma may suggest that the nucleus composition is inhomogeneous (though all molecules were not observed on the same day and we cannot exclude time variations of the outgassing). 
This analysis is consistent with the results of some previous works:
(1) the offsets observed between molecular (HNC, DCN, HDO) and continuum emission  on interferometric maps  made with the Owens Valley Radio Observatory (OVRO) were interpreted as due to  jets enriched in these molecules \citep{blake};
(2)  Monte Carlo simulations suggest that the different molecular jets observed at optical wavelengths (OH, CN, C$_{2}$) originate in several active areas of different chemical compositions \citep{lederer02}.
An explanation would be that the huge nucleus of comet Hale-Bopp is actually made of several pieces that have not undergone the same formation and/or evolution processes. 
This would be in contrast with comet 73P/Schwassmann-Wachmann 3 whose fragments were found similar in chemical properties, suggesting an homogeneous nucleus \citep{dello}.

The results obtained in this paper demonstrate the benefit and uniqueness of millimeter interferometry to investigate nucleus outgassing properties and physical processes acting in the coma. 
Last and not least, by solving the contentious issue regarding the role of nongravitational forces in comet Hale-Bopp orbit, we have shown that this technique is also powerful for astrometry purposes. 
ALMA, with its unprecedented sensitivity, angular resolution, and instantaneous $uv$-coverage, will be a very powerful tool in the future for cometary science. 
  
\begin{acknowledgements}
We acknowledge the IRAM staff for help provided during the observations and for data reduction, and dedicate this paper to J.~E. Wink (1942-2000).
We thank P. Rocher (IMCCE) for providing us the ephemeris of the comet. 
This work has been supported by the Programme national de plan\'etologie (PNP) and by the Action sp\'ecifique ALMA (INSU).
\end{acknowledgements}

\end{document}